\documentstyle[twocolumn,epsf,aps,prb,floats]{revtex}
\begin{document}
\draft

\twocolumn[\hsize\textwidth\columnwidth\hsize\csname @twocolumnfalse\endcsname

\title{Noncrossing Approximation for the Anisotropic Kondo Model:
       Charge Fluctuations in a Quantum Box}
\smallskip
\author{Eran Lebanon, Avraham Schiller, and Vilen Zevin}
\address{
         Racah Institute of Physics, The Hebrew University,
         Jerusalem 91904, Israel}
\date{\today}
\maketitle
\smallskip

\begin{abstract}
The noncrossing approximation (NCA) is generalized to
the multi-channel Kondo-spin Hamiltonian with arbitrary
anisotropic exchange couplings and an external magnetic
field, and applied --- in the framework of Matveev's
mapping --- to the charge fluctuations in a single-electron
box at the Coulomb blockade. The temperature dependences of
the charge step and the capacitance are calculated for a
narrow point contact. At low temperatures and close to the
degeneracy point, the capacitance line shape exhibits
an approximate scaling with $U/\sqrt{T}$, where $U$ is
the deviation in gate voltage from the degeneracy point.
This scaling relation is proposed as a sharp experimental
diagnostic for the non-Fermi-liquid physics of the system
at low temperatures. Both the reliability and
shortcomings of the Kondo NCA are discussed in detail.
Through comparison with poor-man's scaling, we are able
to pinpoint the omission of particle-particle processes
as the origin of the NCA flaws. An extended diagrammatic
scheme is devised to amend the NCA flaws.
\end{abstract}

\smallskip
\pacs{PACS numbers: 72.15.Qm, 73.23.Hk, 75.20.Hr}
\smallskip

]
\narrowtext

\section{Introduction}

The rapid development of nanofabrication techniques has opened new
possibilities for the study of the Kondo effect in mesoscopic
systems.  Using ultra small quantum dots it is now possible to
measure Kondo-assisted tunneling through a single tunable
magnetic impurity.~\cite{QD1,QD2,QD3,QD4,Unitary_limit}
Using scanning tunneling spectroscopy, one can directly probe the local
electronic structure around an isolated magnetic adatom on a metallic
surface.~\cite{STM98_ce,STM98_co,Mirage_00} It is even becoming
experimentally feasible to address the competition between intersite
magnetic locking and Kondo-type correlations in the two-impurity
cluster.~\cite{Co_dimer,GM99} Perhaps most intriguing, though,
are the possible realizations of the two-channel
Kondo effect,~\cite{NB80,Zawadowski80} which is a prototype for a
non-Fermi-liquid ground state in correlated electron systems.~\cite{CZ98}

There are several leading candidates for the observation of the
two-channel Kondo effect in mesoscopic systems. Most notably,
zero-bias anomalies seen in metallic point
contacts,~\cite{RB92,RLvDB94,von_Delft_et_al} which feature
square-root temperature and voltage dependence of the
differential conductance; scaling of the differential conductance
with $eV/k_BT$ ($eV$ being the applied bias); and the absence of
a Zeeman splitting under an applied magnetic field --- all of
which are in line with the notion of two-channel Kondo scattering
off nonmagnetic two-level tunneling systems (TLS).~\cite{CZ98,HKH94}
Two-channel Kondo scattering off TLS was also recently
invoked~\cite{KZ00} to explain the shape and scaling behavior
of the out-of-equilibrium energy distribution function of
quasiparticles in copper and gold wires.~\cite{Pothier_et_al}
In a separate recent experiment on the charging of a semiconductor
quantum dot, Berman {\em et al.}~\cite{Ashoori99} observed capacitance
line shapes in accordance with Matveev's prediction for the
two-channel Kondo screening of quantum charge fluctuations
on the dot.~\cite{Matveev91,Matveev95} The two nearly degenerate
levels in this picture correspond to the two available charge
configurations at the Coulomb blockade.

Common to both the TLS and Coulomb-blockade scenarios for
the two-channel Kondo effect is a strong anisotropy in
the effective spin-exchange interaction.
Specifically, in TLS one generically has~\cite{CZ98}
$J_z \gg J_x > J_y = 0$ ($J_i$ being the $i$th component of the
effective spin-exchange interaction), while $J_z = 0$ and
$J_x = J_y \neq 0$ in the case of the Coulomb
blockade.~\cite{Matveev91}
That the intermediate-coupling fixed point of the two-channel
Kondo model is stable against exchange anisotropy is well
known from numerical renormalization-group~\cite{PC91,ALPC92}
and conformal field theory~\cite{ALPC92,Ye96,comment_on_stability}
studies. However, there is a clear lack in quantitative treatments
that cover all temperature and field regimes of the model, as is
required for each of the two systems mentioned above. This is
especially true of the nonequilibrium scattering off TLS, where
theoretical efforts have focused on the related
${\rm SU}(2)\!\times\!{\rm SU}(2)$ two-channel Anderson
model,~\cite{HKH94} rather than the actual
anisotropic Kondo-spin Hamiltonian appropriate for this case.
We also note a growing interest in quantitative theories of
the single-channel anisotropic Kondo model,~\cite{Gergely}
particularly in connection with the Ohmic dissipative two-state
system,~\cite{DTSS1,DTSS2,DTSS3,DTSS4} and the heavy fermion
compound Ce$_{0.8}$La$_{0.2}$Al$_3$.~\cite{Osborn,Kevin}

In this paper, we extend the noncrossing approximation~\cite{Bickers87}
(NCA) to the Kondo-spin Hamiltonian with arbitrary anisotropic
spin-exchange couplings, and apply it to the charge fluctuations
in a single-electron box at the Coulomb blockade. The NCA was extensively
used in the 80's to study dilute magnetic alloys, especially Ce-
and Yb-based alloys with large orbital degeneracy. Its usefulness
for treating the multi-channel Anderson model was later emphasized
by Cox and Ruckenstein,~\cite{CR93} who noticed that the NCA
pathology hampering the single-channel case~\cite{Bickers87}
actually corresponds to the exact non-Fermi-liquid power laws
and logarithms of the multi-channel Kondo effect.
Compared to exact methods such as the Bethe ansatz and conformal
field theory, the NCA has the crucial advantage that it is not
restricted to idealized models; it could be used to compute
dynamical properties over extended temperature and parameter
regimes; and it has a natural extension to
nonequilibrium.~\cite{HKH94,WM94} Aside from the Anderson
model, the NCA was also applied to the Coqblin-Schrieffer
Hamiltonian~\cite{CS69} in the form of a self-consistent
ladder approximation,~\cite{Maekawa85} yet no direct
formulation of this approach has been devised to date for
the Kondo Hamiltonian.

Generalizing the NCA to the multi-channel Kondo-spin Hamiltonian
with arbitrary spin-exchange and potential-scattering couplings,
a comprehensive analysis of this approach is presented. It is
shown that the Kondo-NCA (KNCA) correctly describes the
non-Fermi-liquid fixed point
of the multi-channel Kondo effect, similar to the NCA formulation
of the multi-channel Anderson model.~\cite{CR93} Furthermore,
there is quantitative agreement with the Bethe ansatz~\cite{SS91}
for the temperature and field dependence of the magnetic
susceptibility of the isotropic two-channel model. At the same
time, the KNCA has several shortcomings, including a
spurious ferromagnetic Kondo effect, and an inaccurate
exponential dependence of the Kondo temperature on the inverse
coupling constants. We are able to pinpoint these shortcomings
of the NCA as due to the omission of particle-particle diagrams,
which also seems to be the reason for the well-known
NCA failure to describe the Fermi-liquid ground state of the
single-channel Kondo effect. As a first step towards amending
this flaw, an extension of the KNCA is constructed,
incorporating particle-particle and particle-hole diagrams on
equal footing.

Applying the KNCA to the charge fluctuations in a quantum box
with a single-mode junction, we propose an approximate scaling
relation for the capacitance line shape as a sharp experimental
diagnostic for the low-temperature, non-Fermi-liquid regime
of the two-channel Kondo effect. This approximate scaling
relation should prove useful in analyzing future experiments,
especially on metallic boxes which are more favorable to
display a fully developed two-channel Kondo effect.~\cite{ZZW00}

The remainder of the paper is organized as follows: 
Section~\ref{sec:mapping} briefly reviews Matveev's mapping
of the charge fluctuations in a single-electron box onto an
anisotropic two-channel Kondo Hamiltonian. The KNCA is then
presented and analyzed in Sec.~\ref{sec:Kondo-NCA}. Technical
details of the KNCA and an extended discussion of aspects
of this approach not directly related to the Coulomb blockade
are deferred to Appendixes~\ref{app:appA} and \ref{app:appB}.
The application of the KNCA to the charge fluctuations
at the Coulomb blockade is reported in turn in
Sec.~\ref{sec:results}, followed by a discussion of the main
results in Sec.~\ref{sec:discussion}. Appendix~\ref{app:appC}
is devoted an extension of the KNCA that properly
preserves particle-hole symmetry.

\section{Mapping the Coulomb blockade onto an anisotropic Kondo problem}
\label{sec:mapping}

The fundamental connection between the charge fluctuations in a
single-electron box at the Coulomb blockade and the Kondo problem
was formulated in a key paper by Matveev.~\cite{Matveev91} Below
we briefly outline Matveev's mapping of the two problems, setting
our notations in doing so.

Consider a single-electron box, connected to a lead by a narrow
point contact. For a sufficiently narrow point contact, there
will be just a single mode weakly connecting the two sides of
the constriction. Such a weak link can be described by the
tunneling Hamiltonian
\begin{equation}
{\cal H}_{T} = \sum_{kp\sigma} \left \{
        t_{kp} c_{k\sigma}^{\dagger} c_{p\sigma} + h.c. \right \} ,
\label{H_tun}
\end{equation}
where $c_{k\sigma}^{\dagger}$ ($c_{p\sigma}^{\dagger}$) creates
a conduction electron with spin projection $\sigma$ in the
lead (box), and $t_{kp}$ are the corresponding tunneling matrix
elements. To simplify our notations, we use the indices $k$ and
$p$ throughout this section to label the single-electron levels
in the lead and in the box, respectively. We further assume that
the level spacing inside the quantum box is sufficiently small
that a continuum-limit
description can be used. As recently emphasized by Zar\'and
{\em et al.},~\cite{ZZW00} it is difficult to realize this
condition in present semiconducting devices, due to the exponentially
small many-body energy scale (Kondo temperature) involved.
We shall return discuss this assumption later on.

In addition to the tunneling Hamiltonian of Eq.~(\ref{H_tun}),
one has the terms describing the isolated lead and box.
While the lead can be modeled by a simple noninteracting band
with dispersion $\epsilon_k$,
\begin{equation}
{\cal H}_{L} = \sum_{k\sigma} \epsilon_k
               c_{k\sigma}^{\dagger}c_{k\sigma} ,
\end{equation}
for the quantum box one has to consider also the charging energy
of the box. Specifically, setting the filled Fermi sea as our
reference state and measuring all single-particle energies
relative to the Fermi level, the charge inside the box is given by
\begin{equation}
Q = -e\sum_{p\sigma}
     [ c_{p\sigma}^{\dagger}c_{p\sigma} - \theta(-\epsilon_p) ] ,
\end{equation}
which has the associated charging energy
\begin{equation}
E_{C} = \frac{Q^2}{2C_B} + \phi Q .
\end{equation}
Here $C_B$ is the capacitance of the quantum box, and $\phi$
is the electrostatic potential in the box measured relative
to the lead. The latter is
related to the applied gate voltage by some geometric
factor. Thus, the full Hamiltonian of the system
has the form ${\cal H} = {\cal H}_{L} + {\cal H}_{B} +
{\cal H}_{T}$, where
\begin{equation}
{\cal H}_{B} = \sum_{p\sigma} \epsilon_p
               c_{p\sigma}^{\dagger}c_{p\sigma} +
               \frac{Q^2}{2C_B} + \phi Q
\end{equation}
describes the isolated single-electron box, which has
the single-particle levels $\epsilon_p$.

\begin{figure}
\centerline{
\vbox{ \epsfxsize=60mm \epsfbox {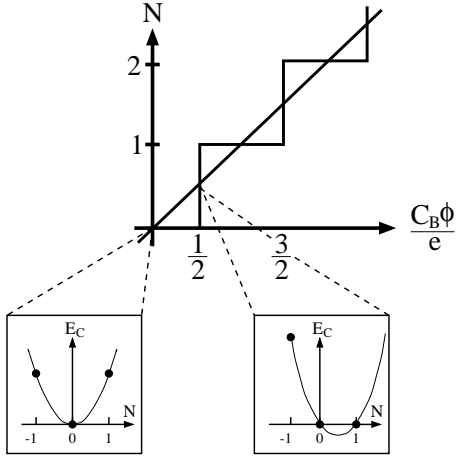}}
}\vspace{10pt}
\caption{Zero-temperature charging of the isolated quantum box,
         in the absence of tunneling to the lead. For a given
	 value of the electrostatic potential $\phi$, the
	 charging energy $E_{C} = (eN)^2/2C_B - e\phi N$ is
	 parabolic in the number of excess electrons in the
	 box, $N = -Q/e$. Each time $\phi C_B/e$ is equal to
	 half an integer, there are two degenerate charge
	 configurations that minimize $E_C$. This results in
	 abrupt jumps in $N$ each time $\phi C_B/e$ crosses
	 a half-integer value.}
\label{fig:fig2}
\end{figure}

For zero tunneling and zero temperature, the charge in the
quantum box is determined by minimizing $E_C$ with respect to
$Q$. Given the discreteness of the charge, this results in
abrupt changes in the ground state of the quantum box each
time $\phi C_B/e$ crosses the half-integer value $N + 1/2$,
corresponding to the transition
from $Q = -eN$ to $Q = -e(N + 1)$ (see Fig.~\ref{fig:fig2}).
Precisely at $\phi C_B/e = N + 1/2$ the two charge
configurations are degenerate, giving rise to strong charge
fluctuations in the box.

The effect of weak tunneling is to smear the sharp charge
steps at $\phi = e(N + \frac{1}{2})/C_B$, yet in a nontrivial manner
to be explored below. To this end, let us focus on the vicinity
of a particular charge step separating $Q = -eN$ from
$Q = -e(N + 1)$. Labeling the deviation from the degeneracy
point by $U = \phi - e(N + \frac{1}{2})/C_B$,
we concentrate on $|U| \ll e/C_B$ and $k_B T \ll e^2/C_B$, such
that all charge configurations other than $Q = -eN$ and
$Q = -e(N + 1)$ are thermally inaccessible.~\cite{comment_on_exp}
One can remove all higher energy charge configurations in
this case by introducing the operators $P_0$ and $P_1$, which
project onto the $Q = -eN$ and $Q = -e(N + 1)$ subspaces,
respectively. Omitting a $U$-independent reference energy,
the resulting low-energy Hamiltonian is written as
\begin{eqnarray}
{\cal H}_{eff} &=& \left(
          \sum_{k\sigma} \epsilon_k c_{k\sigma}^{\dagger} c_{k\sigma} +
          \sum_{p\sigma} \epsilon_p c_{p\sigma}^{\dagger} c_{p\sigma}
          \right)(P_0 + P_1) \nonumber \\
&+& \sum_{kp\sigma} \left\{
          t_{kp} P_0 c_{k\sigma}^{\dagger} c_{p\sigma} P_1 +
          t_{kp}^{\ast} P_1 c_{p\sigma}^{\dagger} c_{k\sigma} P_0
          \right \}
\nonumber \\
&-& \frac{eU}{2} \left( P_1 - P_0 \right) - eU \left (
          N + \frac{1}{2}\right ) \left( P_1 + P_0 \right) .
\label{H_eff}
\end{eqnarray}

To illustrate the connection between Eq.~(\ref{H_eff}) and
the two-channel Kondo Hamiltonian, it is useful to
adopt the following $2\times2$ matrix representation of
the two projection operators:
\begin{equation}
P_0 \longleftrightarrow \left(
\begin{array}{cc}
   0 & 0 \\
   0 & 1
\end{array} \right) , \;\;\;\;\;\;\;\;\;\;
P_1 \longleftrightarrow \left(
\begin{array}{cc}
   1 & 0 \\
   0 & 0
\end{array} \right) .
\end{equation}
In addition, we change our notation for the conduction
electrons by introducing the isospin label
$\alpha = \pm$, distinguishing the quantum-box electrons,
$c^{\dagger}_{k\sigma\alpha=-}$, from the lead electrons,
$c^{\dagger}_{k\sigma\alpha=+}$. Neglecting all momentum
dependence of the tunneling matrix elements, i.e., taking
$t_{kp} = t$, Eq.~(\ref{H_eff}) acquires the form
\begin{eqnarray}
{\cal H}_{eff} &=& \sum_{k\sigma \alpha} \epsilon_{k\alpha}
            c_{k \sigma \alpha}^{\dagger} c_{k \sigma\alpha}
            - eU\!\left( N + \frac{1}{2} \right )
            - \frac{eU}{2}\!\left(
            \begin{array}{cr}
                   1 & 0 \\
                   0 & -1
            \end{array} \right) \nonumber\\
&+& t \sum_{kk'\sigma} \left \{
            c_{k \sigma -}^{\dagger} c_{k' \sigma +}
            \left( \begin{array}{cc}
                   0 & 1 \\
                   0 & 0
            \end{array} \right)
            + h.c. \right \} .
\end{eqnarray}
Finally, expressing the $2 \times 2$ matrices in terms of
a spin-$\frac{1}{2}$ isospin operator $\vec{S}$, our
effective Hamiltonian becomes
\begin{eqnarray}
{\cal H}_{eff} &=& \sum_{k \sigma \alpha} \epsilon_{k}
            c_{k \sigma \alpha}^{\dagger} c_{k \sigma \alpha}
            - eU S_{z} - eU ( N + 1/2 ) \nonumber \\
&+& \frac{t}{2} \sum_{k k' \sigma \alpha \alpha'} \left \{
     c_{k \sigma \alpha}^{\dagger} \sigma_{\alpha\alpha'}^{-}
     c_{k' \sigma \alpha'}S^{+} + h.c. \right \} ,
\label{H_matveev}
\end{eqnarray}
where $\sigma^{\pm} = \sigma_x \pm i\sigma_y$ are Pauli
matrices. Here we have omitted for convenience the isospin
index from the conduction-electron energies in the first
term of Eq.~(\ref{H_matveev}), which amounts to taking a
single joint density of states, $\rho(\epsilon) = \sum_k
\delta( \epsilon - \epsilon_k)$, for the quantum box and the
lead. Since all quantities related to $\vec{S}$ in the problem
depend on the product of the two density of states in the
box and in the lead, $\rho(\epsilon)$ is to be understood
as the geometric average of the two density of
states.~\cite{comment_on_rho}

Up to a constant shift in energy, Eq.~(\ref{H_matveev}) is
the Hamiltonian of an anisotropic two-channel Kondo impurity
in a local magnetic field. Specifically, $J_{\perp} = 2t$
plays the role of the transverse exchange coupling, while
$\mu_B g_J H = eU$ corresponds to an applied magnetic field.
The longitudinal exchange coupling in this mapping is set
equal to zero. Note that the exchange interaction in
Eq.~(\ref{H_matveev}) acts on the isospin index, whereas
the physical spin is reduced to a passive spectator. Hence
the associated many-body screening is that of the charge
fluctuations in the quantum box, not that of a physical spin.
More specific, the charge in the box is related to
the isospin magnetization through
\begin{equation}
\langle Q \rangle = -e \left( N + \frac{1}{2} \right)
        - e \langle S_z \rangle ,
\label{Q_mag}
\end{equation}
while the capacitance of the junction,
$C(U, T) = -\partial \langle Q \rangle/\partial U$, is
equal to $e^2/\left (\mu_B g_J \right)^{2}$ times the isospin
susceptibility,~\cite{comment_on_chi_zz}
\begin{equation}
\chi(H, T) = \mu_B g_J
             \frac{\partial \langle S_z \rangle}{\partial H} .
\label{sus-def}
\end{equation}

As is well known, the low-temperature, low-field properties
of the two-channel Kondo Hamiltonian are governed by an
intermediate-coupling non-Fermi-liquid
fixed point,~\cite{CZ98} which differs markedly from the
infinite-coupling fixed point of the single-channel case. In
particular, the zero-field susceptibility diverges with
decreasing temperature according to 
$\chi \propto (1/T_K) \log(T_K/T)$,
which translates in the present context to an infinite charging
slope at $U = 0$ and $T = 0$. $T_K$ in the above expression is
the Kondo temperature, which marks the crossover from weak coupling.
For the anisotropic Hamiltonian of Eq.~(\ref{H_matveev}) it is
given by~\cite{Matveev91}
\begin{equation}
k_B T_K = (D \rho_0 t)
        \exp \left [ -\frac{\pi}{4\rho_0 t} \right ] ,
\label{T_K_Matveev}
\end{equation}
where $D \sim e^2/C_B$ is the effective conduction-electron bandwidth,
and $\rho_0$ is the geometric average of the lead and the box
density of states at the Fermi level.

Equation~(\ref{T_K_Matveev}) is due to Matveev, who studied the
Hamiltonian of Eq.~(\ref{H_matveev}) within a perturbative
renormalization-group approach.~\cite{Matveev91} Based on
perturbation theory, this approach breaks down as soon as
the renormalized exchange couplings are large, limiting its
applicability to the weak-coupling regime. Our objective is
to develop a quantitative description of all temperature and
field regimes, including the low-temperature, non-Fermi-liquid
regime. To this end, we first extend the noncrossing
approximation to the Kondo Hamiltonian with arbitrary
anisotropic spin-exchange couplings.

\section{Noncrossing Approximation}
\label{sec:Kondo-NCA}

As mentioned above, to date there is no direct formulation of
the NCA for the Kondo-spin Hamiltonian, let alone its strongly
anisotropic version of Eq.~(\ref{H_matveev}). In fact, previous
applications of the NCA to the multi-channel Kondo effect
have been restricted to the ${\rm SU}(N)\!\times\!{\rm SU}(M)$
multi-channel Anderson model.~\cite{CZ98} While the two-channel
Anderson model has often been used to model uranium and
cerium ions in appropriate crystalline-electric-field
environments,~\cite{CZ98} it does not directly apply to the
problem at hand, or to any realization of two-level systems
in a metal. Furthermore, we note that the
anisotropic Kondo Hamiltonian cannot be derived from the
Anderson Hamiltonian via the Schrieffer-Wolff
transformation,~\cite{SW66} which illustrates the
independence of the two models. Below we extend the
formulation of the NCA to the multi-channel Kondo Hamiltonian
with arbitrary spin-exchange and potential-scattering couplings.

As we are interested in devising a general approach,
let us consider in this section the fully anisotropic
$M$-channel Kondo Hamiltonian given by
\begin{eqnarray}
{\cal H} &=& \sum_{k} \sum_{n = 1}^{M} \sum_{\sigma = \uparrow,\downarrow}
             \epsilon_k c^{\dagger}_{kn\sigma} c_{kn\sigma}
             - \mu_B g_J H S_z
\nonumber \\
&+& \sum_{\mu = x, y, z} \frac{J_{\mu}}{2}
             S_{\mu} \sum_{n = 1}^{M}
             \sum_{k k' \sigma \sigma'}
             \sigma^{\mu}_{\sigma \sigma'}
             c^{\dagger}_{kn\sigma} c_{k'n\sigma'} \nonumber\\
&+& \frac{J_0}{4} \sum_{n = 1}^{M} \sum_{\sigma = \uparrow,\downarrow}
             \sum_{k k'} c^{\dagger}_{kn\sigma} c_{k'n\sigma} .
\label{H_multi_channel}
\end{eqnarray}
Here, $\vec{S}$ is the impurity spin; $\sigma^{\mu}$ are Pauli
matrices; $J_x$, $J_y$, and $J_z$ are independent spin-exchange
couplings; $J_0/4$ represents potential scattering at the impurity
site; and $H$ is a local applied magnetic field.
The Hamiltonian of Eq.~(\ref{H_matveev}) corresponds in this
notation to $M = 2$, $J_x = J_y = 2t$, $J_z = J_0 = 0$, and
$\mu_B g_J H = eU$ [note that $\sigma$ labels the channel index
in Eq.~(\ref{H_matveev}), whereas in Eq.~(\ref{H_multi_channel})
it labels the spin index]. The inclusion of potential scattering
in Eq.~(\ref{H_multi_channel}) is mainly intended to allow for
comparison with the Coqblin-Schrieffer Hamiltonian and the
self-consistent ladder approximation of Maekawa
{\em et al.}~\cite{Maekawa85} As we shall see, though, its
inclusion will provide considerable insight into some of the
underlying features of the NCA.

\begin{figure}
\centerline{
\vbox{ \epsfxsize=80mm \epsfbox {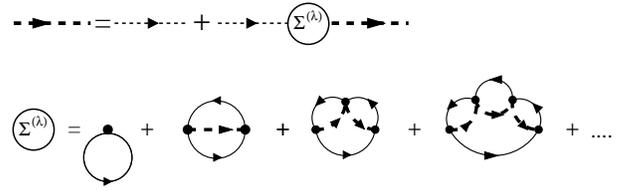}}
}\vspace{10pt}
\caption{Diagrammatic representation of the noncrossing
         approximation for the pseudo-fermion Green function.
	 Here thin (thick) dashed lines represent bare (fully
	 dressed) $f$ Green functions, while full lines stand
	 for bare conduction-electron propagators. In the
	 Coqblin-Schrieffer limit, $J_0 = J_x = J_y = J_z$,
	 these diagrams coincide with the self-consistent
	 ladder approximation of Maekawa
	 {\em et al.},~\cite{Maekawa85} which can also be
	 derived from the NCA equations for the Anderson
         impurity model by taking the appropriate Kondo
	 limit.~\cite{Bickers87}}
\label{fig:fig3}
\end{figure}

We proceed by introducing Abrikosov's slave-fermion 
representation,~\cite{Abrikosov65} which generalizes in the
Anderson model to the slave-boson representation.~\cite{SB_mapping}
In this representation, one assigns a pseudo fermion to each
impurity spin state according to
\begin{eqnarray}
f_{\uparrow}^{\dagger} |0\rangle \longleftrightarrow
             |\uparrow\rangle \; ,\;\;\;\;\;\;
             f_{\downarrow}^{\dagger} |0\rangle \longleftrightarrow
             |\downarrow\rangle.
\end{eqnarray}
The physical subspace corresponds to the constraint ${\hat N}_f
\equiv\sum_{\sigma}f_{\sigma}^{\dagger}f_{\sigma}=1$, which
represents the fact that the impurity spin has only two possible
states. This constraint distinguishes the pseudo fermions from
ordinary fermions.

A convenient way to calculate physical quantities 
is to work within a grand canonical ensemble with respect to
the number of pseudo fermions. These are assigned a chemical
potential $\lambda$, modifying the Hamiltonian from ${\cal H}$
to ${\cal H}-\lambda{\hat N}_f$.
The projection onto the physical subspace is carried out 
by taking the limit $\lambda\rightarrow-\infty$.
Specifically, physical averages are given by
\begin{equation}
\langle \hat{O} \rangle _{phys} = \frac{1}{Z_{imp}}
        \lim_{\lambda \rightarrow -\infty}
        e^{-\beta\lambda} \langle \hat{O} \hat{N}_f \rangle_{\lambda} ,
\label{O_average}
\end{equation}
where
\begin{equation}
Z_{imp} = \lim_{\lambda \rightarrow -\infty} e^{-\beta\lambda}
          \langle{\hat N}_f\rangle_{\lambda}
\label{Z_imp}
\end{equation}
is the ``impurity contribution'' to the partition function. Here
subscripts $\lambda$ denote averages within the grand-canonical
ensemble. In practice, one can drop the $\hat{N}_f$ operator in
Eq.~(\ref{O_average}) for those physical operators $\hat{O}$
that give zero when acting on the $\hat{N}_f = 0$ subspace,
which greatly simplifies the calculations.

A central quantity in the calculation of physical observables
is the pseudo-fermion Green function, $G_{\sigma}^{(\lambda)}(z)$,
which enters in its projected form
$G_{\sigma}(z) = \lim_{\lambda \rightarrow -\infty}
G_{\sigma}^{(\lambda)}(z-\lambda)$. The latter will
be referred to hereafter as the pseudo-fermion Green
function.~\cite{comment_on_sigma} The
pseudo-fermion Green function takes the standard form
$G_{\sigma}(z)=[z- \epsilon_{\sigma} - \Sigma_{\sigma}(z)]^{-1}$,
where $\Sigma_{\sigma}(z)$ is the projected self-energy, and
$\epsilon_{\sigma} = -\sigma \frac{1}{2} \mu_B g_J H$ is the bare
energy of the impurity spin (we use $\sigma=\uparrow,\downarrow$
and $\sigma = \pm$ interchangeably to label the spin $z$
component). The Kondo-NCA (KNCA) is defined by a particular
set of self-energy diagrams, shown in Fig.~\ref{fig:fig3}.
Although additional diagrams without any crossings do exist
for the Kondo-spin Hamiltonian (specifically the
particle-particle ladder, see Appendix~\ref{app:appC}),
we refer to this approximate scheme as the NCA, because
it has the same general structure as that of the conventional
NCA for the Anderson Hamiltonian.

Working within the slave-fermion representation, we write the 
interaction term of Eq.~(\ref{H_multi_channel}) in the form
\begin{equation}
\sum_{n=1}^{M} \sum_{kk'} \sum_{\alpha\beta\alpha'\beta'}
            V_{(\alpha\beta)(\alpha' \beta')} c_{nk\alpha}^{\dagger}
            c_{nk'\alpha'} f_{\beta'}^{\dagger} f_{\beta} ,
\label{V_abgd_1}
\end{equation}
where 
\begin{equation}
V_{(\alpha\beta)(\alpha'\beta')} = \sum_{\mu=0, x, y, z}
            \frac{J_{\mu}}{4} \sigma^{\mu}_{\alpha\alpha'}
            \sigma^{\mu}_{\beta'\beta} .
\label{V_abgd_2}
\end{equation}
Here $\sigma^0$ denotes the $2\times2$ unity matrix. Viewing the
$V_{(\alpha\beta)(\alpha'\beta')}$ coefficients as the elements
of a $4\times4$ matrix, this choice of labels allows for a convenient
summation of the ladder diagram of Fig.~\ref{fig:fig3}. Details
of the derivation are provided in Appendix~\ref{app:appA}. For a
zero magnetic field, when both spin orientations are equivalent,
one obtains~\cite{comment_on_constant_part}
\begin{equation}
\Sigma(\epsilon + i\delta) = \frac{M}{2} \sum_k f(-\epsilon_k)
               \Delta(\epsilon - \epsilon_k + i\delta) ,
\label{sigma_H=0}
\end{equation}
where
\begin{equation}
\Delta(\epsilon + i\delta) = -\sum_{\mu=0, x, y, z}
                    \frac{A_{\mu}}{1 - A_{\mu}\Pi(\epsilon + i\delta)} ,
\label{Delta_def}
\end{equation}
\begin{equation}
\Pi(\epsilon + i\delta) =
                  - \sum_k f(\epsilon_k) G(\epsilon+\epsilon_k+i\delta) .
\label{Pi_H=0}
\end{equation}
Here $f(\epsilon)$ is the Fermi-Dirac distribution function,
and the $A_{\mu}$ coefficients are defined as
\begin{eqnarray}
A_0 &=& \frac{1}{4} \left (J_0 + J_x + J_y + J_z \right) ,
\label{A_0_def}\\
A_i &=& \frac{1}{4} \left (J_0 + 2J_i - J_x - J_y - J_z \right)
\label{A_i_def}
\end{eqnarray}
($i = x, y, z$). For
$J_0 = J_x = J_y = J_z$, corresponding to the Coqblin-Schrieffer
Hamiltonian, Eqs.~(\ref{sigma_H=0})--(\ref{Pi_H=0}) coincide
with the self-consistent ladder approximation of Maekawa
{\em et al.}~\cite{Maekawa85}

\begin{figure}
\centerline{
\vbox{\epsfxsize=85mm \epsfbox {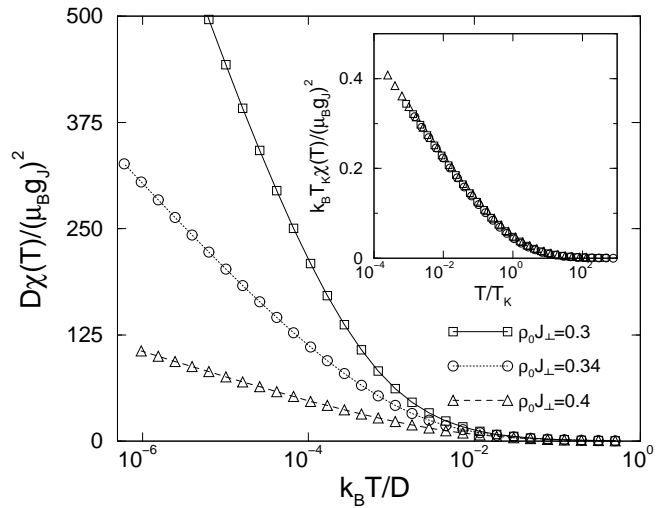}}
}\vspace{10pt}
\caption{KNCA results for the zero-field magnetic
         susceptibility of the anisotropic two-channel
	 Kondo impurity with $J_z = 0$ and different
         $\rho_0 J_{\perp}$. Here and throughout the
	 paper we use a semi-circular conduction-electron
	 density of states with half-width $D$. At low
	 temperatures, $\chi(T)$ diverges logarithmically
	 with decreasing temperature, in accordance with
	 exact results. The robustness and universality of
	 the logarithmic temperature dependence is demonstrated
	 in the inset, where all lines have been collapsed onto
	 a single curve by rescaling $\chi(T)$ and $T$
	 with the Kondo temperature, $T_K$. The
	 latter is extracted according to
	 Eq.~(\ref{Bethe_an_eq}), taking the values
         $k_B T_K/D = 4\!\times\!10^{-4}, 1.15\!\times\!10^{-3}$,
         and $3.85\!\times\!10^{-3}$ for
	 $\rho_0 J_{\perp} = 0.3, 0.34$, and $0.4$,
	 respectively.}
\label{fig:fig4}
\end{figure}

As stated above, Eqs.~(\ref{sigma_H=0})--(\ref{Pi_H=0})
are for a zero magnetic field. However, to study the smearing
of the charge step for the quantum box, it is necessary to
include also a local magnetic field. The addition of a finite
magnetic field clearly distinguishes the present NCA
formulation from those of the Anderson~\cite{Bickers87}
and Coqblin-Schrieffer~\cite{Maekawa85} Hamiltonians.
While in the latter two cases one simply replaces the
zero-field pseudo-fermion propagators entering the NCA
equations with spin-dependent ones, the structure of
Eqs.~(\ref{sigma_H=0})--(\ref{Pi_H=0}) is substantially
modified once $H$ is nonzero. The generalization of
Eqs.~(\ref{sigma_H=0})--(\ref{Pi_H=0})
to a nonzero field is specified in Appendix~\ref{app:appA},
Eqs.~(\ref{Pi_bubble}), (\ref{Sigma_general}), and (\ref{Delta_general}).

Although somewhat cumbersome, the magnetic-field equations
are essential to our discussion. They are solved numerically
by iterations similar to the zero-field ones. From their solution
one can compute the field-dependent magnetization $M(H, T)$,
from which $\chi(H, T)$ follows by direct numerical
differentiation with respect to $H$. Interestingly, we found
it advantageous to use this finite-field differentiation procedure
also when computing the zero-field susceptibility. This somewhat
surprising result stems from the fact that the standard
magnetization bubble acquires a nontrivial vertex correction
that is absent for both the Anderson and Coqblin-Schrieffer
Hamiltonians, and which contributes an important log-$T$ diverging
term to the low-temperature susceptibility of the two-channel
case (see Appendix~\ref{app:appA}, Figs.~\ref{fig:app2} and
\ref{fig:app3}). Deferring
all details of the numerical procedure to Appendix~\ref{app:appA},
below we summarize the main features of the KNCA approach.
All results presented below were obtained for a semi-circular
conduction-electron density of states with half-width $D$:
$\rho(\epsilon) = \rho_0\sqrt{1 - (\epsilon/D)^2}$, where $\rho_0$
is the conduction-electron density of states at the Fermi level.

The KNCA approach presented above has several notable
successes. Taking the zero-temperature limit and performing
a M\"uller-Hartmann~\cite{MH84} type of analysis, one can show
that Eqs.~(\ref{sigma_H=0})--(\ref{Pi_H=0}) reproduce the exact
critical exponents of the multi-channel ($M > 1$) Kondo effect
for antiferromagnetic couplings, similar to the NCA treatment
of the multi-channel Anderson model.~\cite{CR93} Focusing on
antiferromagnetic couplings, the low-temperature susceptibility
for the $M=2$, two-channel case diverges logarithmically with
decreasing temperature, in accordance with exact results.~\cite{CZ98}
For the anisotropic limit of Eq.~(\ref{H_matveev}), this is
demonstrated in Fig.~\ref{fig:fig4}, for different values of
$\rho_0 J_{\perp}$ ($J_{\perp}$ being the transverse spin-exchange
coupling: $J_{\perp} \equiv J_x = J_y$). Most significantly,
there is quantitative agreement with the Bethe ansatz for
the field and temperature dependence of the susceptibility in
the isotropic limit,~\cite{SS91} as shown in Fig.~\ref{fig:fig5}.
Here and throughout the paper we extract the two-channel Kondo
temperature from the Bethe ansatz expression for the slope of the
log-$T$ diverging term in the zero-field susceptibility:~\cite{SS89}
\begin{equation}
\chi(T) \sim \frac{(\mu_B g_J)^2}{20 k_B T_K}\ln(T_K/T) .
\label{Bethe_an_eq}
\end{equation}
Hence the KNCA correctly describes the non-Fermi-liquid
fixed point of the multi-channel Kondo effect both qualitatively
and quantitatively. Finally, Eqs.~(\ref{sigma_H=0})--(\ref{Pi_H=0})
correctly
give rise to a quantum phase boundary between ``antiferromagnetic''
($J_z > -|J_{\perp}|$) and ``ferromagnetic'' ($J_z \le -|J_{\perp}|$)
couplings, each of which is characterized by different threshold
exponents.~\cite{MH84,comment_on_ferro}

At the same time, Eqs.~(\ref{sigma_H=0})--(\ref{Pi_H=0}) have
several clear shortcomings:
\begin{figure}
\centerline{
\vbox{\epsfxsize=85mm \epsfbox {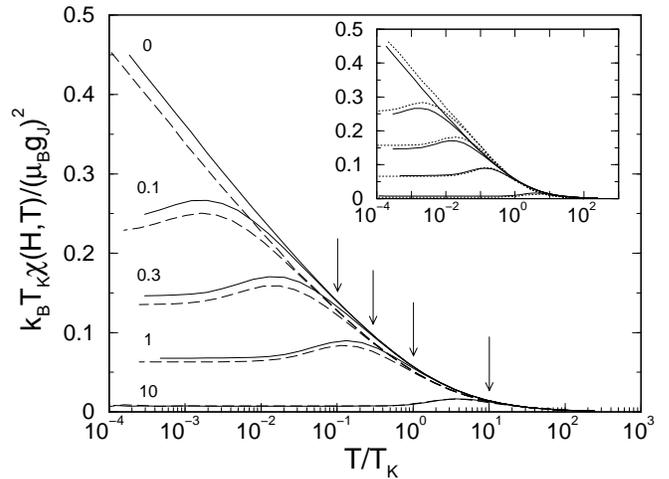}}
}\vspace{10pt}
\caption{Temperature and field dependence of the magnetic
         susceptibility within the KNCA: full lines are
	 for the isotropic two-channel Kondo model with
	 $J_0 = 0$ and $\rho_0 J = 0.2667$
	 ($k_B T_K/D = 2\!\times\!10^{-3}$); dashed lines
	 are for the anisotropic two-channel Kondo model
	 with $J_0 = J_z = 0$ and $\rho_0 J_{\perp} = 0.4$
         ($k_B T_K/D = 3.85\!\times\!10^{-3}$).
	 The applied magnetic field takes the values of
         $\mu_B g_J H/k_B T_K = 0, 0.1, 0.3, 1$, and $10$,
	 according to the labels marking each set of curves.
	 For each field, the associated temperature
	 $T = \mu_B g_J H/k_B$ is indicated by an arrow.
	 Inset: Comparison between the Bethe ansatz (dotted
	 lines, taken from Ref.~\protect\onlinecite{SS91})
         and the KNCA (full lines), for the isotropic
	 two-channel Kondo model. Here $\rho_0 J = 0.2667$
         for the KNCA curves. The values of $\mu_B g_J H/k_B T_K$
         used and their line assignments are identical to
	 those in the main figure.}
\label{fig:fig5}
\end{figure}
(i) A spurious Kondo-type effect (i.e., a low-temperature resonance
with an underlying nonperturbative energy scale) is found for
ferromagnetic couplings, albeit with different characteristic
exponents than in the antiferromagnetic case. For $J_0 > 0$,
this spurious Kondo-type effect extends also to pure potential
scattering;
(ii) Starting from a particle-hole symmetric Hamiltonian, the
KNCA breaks particle-hole symmetry, which is manifest,
for example, in the asymmetric energy dependence of the
conduction-electron $T$-matrix [See Appendix~\ref{app:appA},
Fig.~\ref{fig:app4}];
(iii) Inaccuracies are found in the exponential dependence of Kondo
temperature on the inverse coupling constants. Specifically, the
exponential dependence of $T_K$ on $1/J_{\mu}$ can be estimated
from the location of the lowest pole in the first iteration for
$\Delta(\epsilon)$ at zero temperature. Subsequent iterations
only modify the pre-exponential factor. For antiferromagnetic
couplings with $J_{\perp} > 0$ one thus obtains
\begin{equation}
T_K^{(NCA)} \sim D \exp \left[-\frac{1}{\rho_0 A_0} \right],
\label{T_K_NCA}
\end{equation}
which reduces in the isotropic case [when
$A_0 = 3\rho_0 J/4$, see Eq.~(\ref{A_0_def})] to
$T_K^{(NCA)} \sim D \exp [ -4/3\rho_0 J]$. Here $D$ is the
conduction-electron bandwidth. This should to be contrasted
with the well-known result $T_K \sim D \exp [ -1/\rho_0 J]$,
valid for any number of channels $M$.~\cite{Yosida66} Similarly,
for the anisotropic Hamiltonian of Eq.~(\ref{H_matveev}) one obtains
$T_K^{(NCA)} \sim D \exp [ -2/\rho_0 J_{\perp}]$, compared to
$T_K \sim D \exp[-\pi/2\rho_0 J_{\perp}]$ of Eq.~(\ref{T_K_Matveev}).
Only in the Coqblin-Schrieffer limit, when a nonzero potential
scattering $J_0 = J$ is included, does one recover the correct
exponential dependence of $T_K$ on $1/J$ for
isotropic spin-exchange couplings.

As discussed at length in Appendix~\ref{app:appB}, we are able to
pinpoint these shortcomings of the NCA as due to the omission
of particle-particle diagrams. While particle-particle
diagrams are of higher order in $1/N$ in a formal large-$N$
expansion as per the Anderson Hamiltonian,~\cite{Bickers87}
they must be treated on equal footing in the spin-$\frac{1}{2}$
Kondo model. The omission of particle-particle diagrams leads to
the spontaneous generation of potential scattering, drives the
system to a ferromagnetic Kondo effect, and spoils the exponential
dependence of the Kondo temperature on the inverse coupling
constants. It is our belief that their omission is also responsible
for the well-known NCA failure to describe the Fermi-liquid
ground state of the single-channel Kondo effect. We expand
on these ideas in Appendix~\ref{app:appC}, where an extension
of the KNCA scheme is proposed based on a summation
of all parquet diagrams. It is shown that even the crudest
treatment of the extended scheme gives encouraging results
for the amendment of the above NCA flaws.

It is important to note, though, that since $T_K$ is exponentially
sensitive to the microscopic parameters of the system, it is best
extracted directly from the experiment. Thus, viewing $T_K$ as an
experimental parameter, one can use the KNCA to reliably
study the delicate interplay between the temperature $T$, the
local magnetic field $H$, and the Kondo temperature $T_K$ in
the non-Fermi-liquid regime of the two-channel Kondo effect.

\section{Application of the KNCA to the Coulomb blockade}
\label{sec:results}

In this section, we apply the KNCA to study the smearing of
the charge step for a quantum box, in the case of a narrow
point contact.
Smearing of the charge step was previously studied by various
authors using different approaches. These include
renormalization-group techniques,~\cite{Matveev91} perturbation
theory in the tunneling conductance,~\cite{Grabert94} infinite
summations of subclasses of diagrams,~\cite{GZ94,SS94} and
Monte Carlo simulations.~\cite{Monte_Carlo} However, these
approaches generally fail at low temperatures near the
degeneracy point. Here we exploit the KNCA to reliably
study this regime for a single-mode junction.

As discussed
in Sec.~\ref{sec:mapping}, the excess charge in the quantum
box is related within the two-channel Kondo mapping to the
isospin magnetization [see Eq.~(\ref{Q_mag})],
while the capacitance corresponds to the isospin susceptibility.
Our task therefore amounts to computing the field and temperature
dependence of the magnetization and susceptibility of the
anisotropic two-channel Kondo Hamiltonian with $J_z = 0$.

\subsection{Evolution of the charge step}

Generally speaking, there are three energy scales that govern the
shape of the charge step for the quantum box: $k_B T$, $k_B T_K$,
and $e^2/C_B$. At high temperatures, $k_B T \gg e^2/C_B$, the
classical picture is recovered. Thermal fluctuations are
sufficiently strong to wash out the Coulomb blockade, and a
linear $Q$ vs. $\phi$ curve is obtained with a slope of
$C_B$. The Coulomb-blockade picture emerge as
$k_B T$ is lowered below $e^2/C_B$. Specifically, for
$k_B T \ll e^2/C_B$ there are extended, nearly horizontal charge
plateaus, separated by rather narrow charge steps. The charge plateaus
extend over a length of $\Delta (eU)_{\rm plateau} \approx e/C_B$,
while the charge steps are confined to a width of
$\Delta (eU)_{\rm step} \propto {\rm max}
\{k_B T, k_B T_K\}$.~\cite{comment_on_smearing}
It is only in this temperature regime and in the vicinity of
the charge steps that the mapping onto the Hamiltonian of
Eq.~(\ref{H_matveev}) applies.

\begin{figure}
\centerline{
\vbox{\epsfxsize=85mm \epsfbox {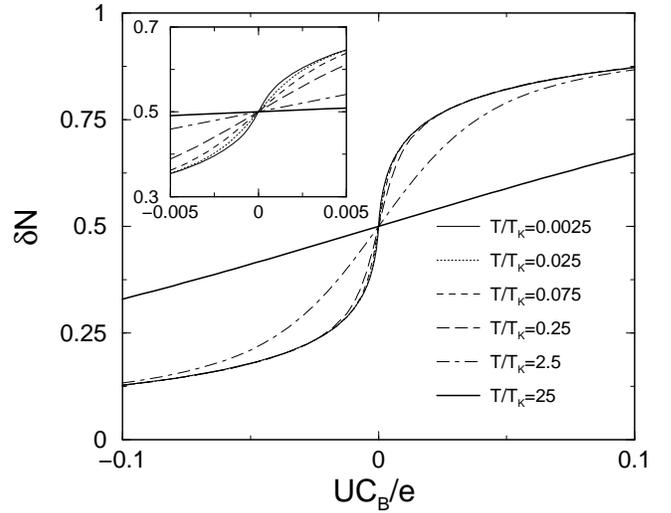}}
}\vspace{10pt}
\caption{Temperature dependence of the charge step
         between the $\langle Q \rangle = -eN$ and
         $\langle Q \rangle = -e(N + 1)$ charge plateaus,
	 as obtained from the KNCA. Here we parameterize the
         transition between the plateaus by
	 $\delta N = - \langle Q \rangle /e - N$.
         The tunneling matrix element is taken to be
	 $\rho_0 t = 0.2$, corresponding to
	 $\rho_0 J_{\perp} = 0.4$ in the anisotropic
         two-channel Kondo representation of
	 Eq.~(\ref{H_multi_channel}). The corresponding
	 Kondo temperature is equal to $k_B T_K/D =
         3.85\!\times\!10^{-3}$. Following
	 Matveev,~\protect\cite{Matveev91} we identify
	 the conduction-electron bandwidth $D$ with $e^2/C_B$,
         such that $eU/D$ is equal to $UC_B/e$. The
	 overall shape of the charge step is mostly
	 unchanged for $T \ll T_K$, however the slope
         around $U = 0$ continues to steepen with
	 decreasing temperature (see inset). Eventually,
	 an infinite slope develops at $U = 0$ as $T \to 0$.}
\label{fig:fig6}
\end{figure}

\begin{figure}
\centerline{
\vbox{\epsfxsize=80mm \epsfbox {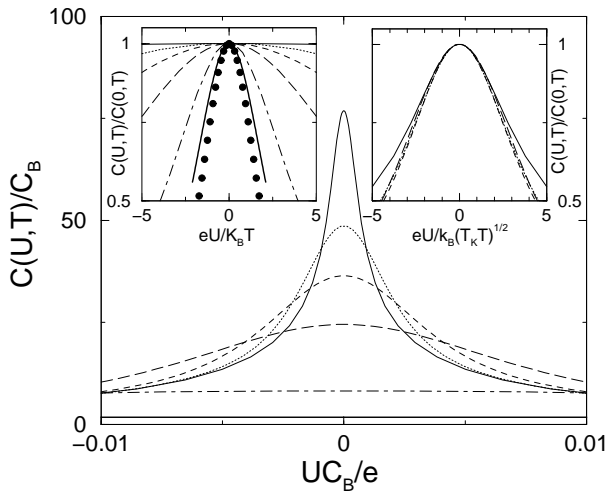}}
}\vspace{10pt}
\caption{Temperature dependence of the capacitance
	 line shape $C(U, T)$. Here the temperatures
         and line-type assignments are the same as
	 in Fig.~\protect\ref{fig:fig6}. With decreasing
         temperature, $C(U, T)$ develops a sharp peak
	 about the center of the charge step ($U = 0$),
	 the height of which diverges logarithmically
         with decreasing $T$. Left inset: The reduced
         capacitance, $C(U, T)/C(0, T)$, plotted versus
	 $x_1 = eU/k_B T$. With increasing temperature,
	 the curves approach the shape of the derivative
	 of the Fermi-Dirac distribution function, normalized
         to unity peak height [i.e., $4e^{x_1}/(1+e^{x_1})^2$,
	 full circles]. Right inset: $C(U, T)/C(0, T)$ plotted
	 versus $x_2 = eU/k_B \sqrt{T T_K}$, for the four
	 temperatures below $T_K$ (i.e., $T/T_K = 0.0025,
         0.025, 0.075$, and $0.25$). For small to intermediate
	 $x_2$, all lines approximately collapse onto a single
	 curve.} 
\label{fig:fig7}
\end{figure}

Restricting attention to $k_B T \ll e^2/C_B$, Fig.~\ref{fig:fig6}
depicts the evolution of the charge step with decreasing $T/T_K$.
For $T_K \ll T$, the charge step is governed by thermal
fluctuations. This is best seen in the capacitance line shape,
which approaches the shape of the derivative of the
Fermi-Dirac distribution function for elevated temperatures
(see Fig.~\ref{fig:fig7}, left inset). For $T_K \ll T$, the
width and shape of the charge step are therefore set by the
temperature. By contrast, quantum fluctuations dominate for
$T \ll T_K$. The overall
extent of the step is proportional to $k_B T_K$, and its shape is
mostly the same as at $T = 0$. Note, however, that the slope at $U = 0$
continues to steepen with decreasing temperature (see inset of
Fig.~\ref{fig:fig6}), approaching an infinite slope as
$T \to 0$.~\cite{Matveev91}
This lack of saturation in the capacitance at $U = 0$ corresponds to
the divergence of the linear susceptibility within the two-channel
Kondo model, which is a hallmark of the non-Fermi-liquid ground
state that develops in this case.

The evolution of the charge step is best resolved in the capacitance
line shape $C(U, T)$, shown in Fig.~\ref{fig:fig7}. When plotted as
a function of $UC_B/e$ (corresponding to $eU/D$ with $D = e^2/C_B$), the
capacitance develops a sharp peak about the center of the charge step
($U = 0$), the height of which diverges logarithmically with decreasing
$T$. To maintain a total integrated weight of one unit charge, there
is a redistribution of weight from the shoulders of the peak to its
center as $T$ is decreased.

The crossover from thermal fluctuations to quantum fluctuations
can be quantified by considering the reduced capacitance,
$C(U, T)/C(0, T)$. To this end, we have plotted the reduced
capacitance of Fig.~\ref{fig:fig7} once versus $x_1 = eU/k_B T$
(left inset), and once verses $x_2 = eU/k_B \sqrt{T_K T}$
(right inset). Clearly, two different qualitative behaviors
are found, depending on whether $T \gg T_K$ or $T \ll T_K$.
For $T \gg T_K$, the reduced capacitance approaches the curve
$4e^{x_1}/(1+e^{x_1})^2$, which is just the derivative of
the Fermi-Dirac distribution function, normalized to unity
peak height. This is to be expected of thermal fluctuations,
when the occupations of the two available charge configurations
in the quantum box follow the Fermi-Dirac distribution
function. A different qualitative behavior is seen for
$T \ll T_K$. Here there is a strong fanning out of the
reduced capacitance when plotted versus $x_1$, but an
apparent scaling when plotted versus $x_2$. As argued below,
this approximate scaling with $x_2$ is an important
characteristic of the non-Fermi-liquid regime of the
two-channel Kondo effect.

\subsection{Approximate scaling with $U/\sqrt{T}$}

To understand the apparent scaling of $U$ with $\sqrt{T}$
for $T \ll T_K$, we go back to the two-channel Kondo
representation of the Coulomb blockade. As is well known, the
intermediate-coupling non-Fermi-liquid fixed point of the
two-channel Kondo model is unstable against an applied
magnetic field, which drives the system to a phase-shifted
Fermi-liquid ground state.~\cite{comment_on_FL} For a weak
magnetic field, $\mu_B g_J |H| \ll T_K$,
this crossover from non-Fermi-liquid to Fermi-liquid
behavior is associated with an energy scale
\begin{equation}
k_B T_x = \frac{(\mu_B g_J H)^2}{k_B T_K} ,
\end{equation}
that depends quadratically on $H$.
Loosely speaking, for $T_x \ll T \ll T_K$ a magnetic field has
only a minor effect on the low-temperature thermodynamics, while
for $T \ll T_x \ll T_K$ the effect of a temperature is small.
In particular, the susceptibility $\chi(H, T)$ is weakly
field dependent for $T_x \ll T$, and is logarithmically
field dependent for $T \ll T_x$.
The connection with the reduced-capacitance plot of
Fig.~\ref{fig:fig7} is made by recalling that $eU$ and
$C(U, T)/C(0, T)$ in the Coulomb blockade map onto
$\mu_B g_J H$ and $\chi(H, T)/\chi(0, T)$, respectively,
in the two-channel Kondo representation. Thus, when plotted
versus $x_2 = \pm \sqrt{T_x/T}$, the reduced capacitance shows a
characteristic crossover from approximately one for $|x_2| < 1$,
to logarithmic in $|x_2|$ for $|x_2| > 1$.

Although the right inset of Fig.~\ref{fig:fig7} seems to suggest
an exact scaling form for the reduced capacitance as a function
of $x_2$ for $T \ll T_K$, we emphasize that this scaling is only
approximate. Namely, writing the reduced capacitance in the form
\begin{equation}
F(x_2, T) = C(U, T)/C(0, T)
\label{F_scaling}
\end{equation}
with $x_2 = eU/k_B \sqrt{T_K T}$, $F$ does not reduce in the
limit $T \ll T_K$ to a function of $x_2$ alone. Nevertheless, the
residual $T$ dependence is logarithmic in nature (see below),
and therefore relatively weak.

So far our discussion has been limited to the KNCA, and to
the $J_z = 0$ limit of the anisotropic two-channel Kondo
model. However, we expect the approximate scaling function
of Eq.~(\ref{F_scaling}) to be generic to all antiferromagnetic
couplings of the anisotropic two-channel Kondo
Hamiltonian. To support this conjecture and to gain
analytical insight into the structure of $F(x_2, T)$,
we consider below the opposite limit of $J_z \gg J_{\perp}$,
for which an exact analytic solution of the two-channel
Kondo model exists for a particular value of $J_z$. At this
special point, known as the Emery-Kivelson point,~\cite{EK92}
one can compute $F(x_2, T)$ exactly, without resorting to the
KNCA. The results for this special limit are summarized in
Fig.~\ref{fig:fig8} and Eqs.~(\ref{EK-1})--(\ref{EK-3}) below.

\subsection{Comparison with the Emery-Kivelson point}

\begin{figure}
\centerline{
\vbox{\epsfxsize=85mm \epsfbox {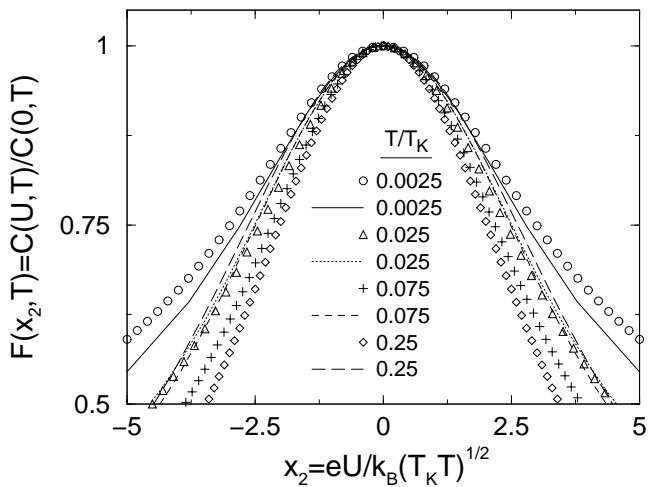}}
}\vspace{10pt}
\caption{Comparison of the proposed scaling function
         $F(x_2, T)$ at the Emery-Kivelson point with the one
         obtained at the Coulomb blockade using the KNCA
         (curves taken from the right inset of Fig.~\ref{fig:fig7}).
         Here symbols correspond to the Emery-Kivelson point,
         while lines mark the curves of the KNCA. Note
         that $T$ is varied over two decades of temperature,
         from $T/T_K = 0.25$ down to $T/T_K = 0.0025$.}
\label{fig:fig8}
\end{figure} 

As noted by Emery and Kivelson,~\cite{EK92} the two-channel
Kondo model with $\rho_0 J_{\perp} \ll \rho_0 J_z = 1$
($J_z = \pi v_F$ in the notations of Ref.~\onlinecite{EK92}) can be
mapped exactly onto a noninteracting ``Majorana'' resonant-level
model, which is the analog of the Toulouse limit for the
single-channel case.~\cite{Toulouse70} At this special point,
one has an exact analytic solution of the anisotropic two-channel
Kondo model, which captures the essential low-temperature physics
of the model. In particular, one can use the Emery-Kivelson point
to extract universal low-temperature behaviors of the two-channel
Kondo model.~\cite{SG95}

Using the exact solution of Emery and Kivelson, we computed
the local-field and temperature dependence of the magnetization
at the solvable point. The resulting expression reads
\begin{equation}
M(H, T) = \frac{(\mu_B g_J)^2 H}{20 T_K} {\rm Re} \left \{
          \frac{\psi(z_{+}) - \psi(z_{-})}{\Delta} \right \} ,
\label{M_EK}
\end{equation}
where $\psi(z)$ is the digamma function,~\cite{abramowitz}
$z_{\pm}$ are equal to
\begin{equation}
z_{\pm} = \frac{1}{2} + \frac{5 T_K}{\pi^2 T} \left (1 \pm \Delta \right) ,
\end{equation}
and $\Delta$ is given by
\begin{equation}
\Delta = \sqrt{ 1 - \left( \frac{\pi \mu_B g_J H}{10 T_K} \right)^2 } .
\end{equation}
Here $T_K$ is the Kondo temperature, defined according to the Bethe
ansatz expression for the slope of the log-$T$ diverging term
in the zero-field susceptibility, Eq.~(\ref{Bethe_an_eq}). With
this definition, $T_K$ is related to the half-width $\Gamma$ of
the Majorana resonant
level (see Ref.~\onlinecite{EK92}) through $T_K = \pi \Gamma/20$.
The field-dependent susceptibility $\chi(H, T)$
is readily obtained from Eq.~(\ref{M_EK}) by differentiating
$M(H, T)$ with respect to $H$.

Focusing on $T < T_K$ and substituting $C(U, T)/C(0, T)$
and $eU$ for $\chi(H, T)/\chi(0, T)$ and $\mu_B g_J H$,
respectively, Fig.~\ref{fig:fig8} compares the proposed
scaling function $F(x_2, T)$ at the solvable point with the
one obtained within the KNCA for the Coulomb blockade.
There is good overall agreement between the two cases, even though
one is dealing with two opposite limits of the anisotropic
two-channel Kondo model, and with two entirely different
methods of calculation (KNCA versus an exact solution based
on mapping onto a ``Majorana'' resonant-level model).
Although the residual $T$ dependence
is more pronounced at solvable point, it is still quite
small considering that $T$ is varied in Fig.~\ref{fig:fig8}
over two decades of temperature.

The great advantage of the solvable point, though, lies in the
explicit form of $F(x_2, T)$, which allows for an analytic study
of the limit $k_B T, e|U| \ll k_B T_K$. Explicitly, omitting terms
that are either a factor of ${\cal O} \left ( (eU/T_K)^2 \right)$
or a factor of ${\cal O} \left ( T/T_K \right)$ smaller than
the leading-order ones, $F(x_2, T)$ is expressed as
\begin{equation}
F(x_2, T) = 1 - \frac{g(x_2)} {h(T)} ,
\label{EK-1}
\end{equation}
where
\begin{equation}
g(x_2) = \psi\!\left( \frac{1}{2} + \frac{x_2^2}{40} \right)
       + \frac{x_2^2}{20}\psi^{(1)}\!\left(
                      \frac{1}{2} + \frac{x_2^2}{40} \right)
       - \psi\!\left (\frac{1}{2} \right)
\label{eq_for_gx2}
\end{equation}
and
\begin{equation}
h(T) = \psi\!\left( \frac{1}{2} + \frac{10 T_K}{\pi^2 T} \right)
            - \psi\!\left (\frac{1}{2} \right) \approx
        \log \left(\frac{7.2 T_K}{T} \right) .
\label{EK-3}
\end{equation}
Here $\psi^{(1)}(z) = d\psi/dz$ is the trigamma
function,~\cite{abramowitz} while $g(x_2) \approx
3 \pi^2 x_2^2/80$ for $|x_2| \ll 1$.
Clearly, $T$ enters $F(x_2, T)$
only logarithmically through $h(T)$, which explains the
rather weak temperature dependence of the proposed scaling
function. We are therefore led to conclude that a scaling plot
of the reduced capacitance versus $x_2 = eU/k_B \sqrt{T_K T}$
provides a sharp diagnostic for the two-channel Kondo scenario,
directly testing the associated non-Fermi-liquid scaling
of $U$ with $\sqrt{T}$ at low temperatures.

\begin{figure}
\centerline{
\vbox{\epsfxsize=85mm \epsfbox {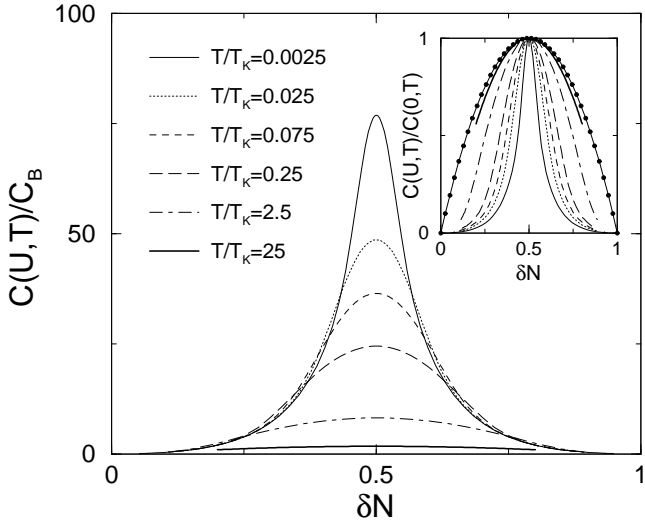}}
}\vspace{10pt}
\caption{The capacitance line shapes of Fig.~\protect\ref{fig:fig7},
         re-plotted as a function of
         $\delta N = - \langle Q \rangle /e - N$. Within the
	 two-channel Kondo representation, this corresponds to
	 plotting the magnetic susceptibility as a function of the
         magnetization. For $T < T_K$, there are strong deviations
	 from the parabolic form of Eq.~(\ref{parabolic}), which
	 characterizes pure thermal fluctuations.
	 Inset: A plot of the reduced capacitance $C(U, T)/C(0, T)$
	 versus $\delta N$. In accordance with predominant thermal
	 fluctuations, for $T \gg T_K$ the reduced capacitance
	 approaches the parabola $4 \delta N (1 - \delta N)$
	 (line with full circles).}
\label{fig:fig9}
\end{figure} 

\subsection{Capacitance versus average charge}

Finally, a different manifestation of the onset of quantum
fluctuations is provided by the inter-relation between the
capacitance line shape and the average charge in the quantum
box. Specifically, in Fig.~\ref{fig:fig9} we have
re-plotted the capacitance line shape of Fig.~\ref{fig:fig7}
as a function of the average number of excess electrons in the
box, parameterized by $\delta N = - \langle Q \rangle /e - N$.
Within the two-channel Kondo representation, this corresponds
to plotting the magnetic susceptibility as a function of the
magnetization. In the case of pure thermal fluctuations, when
$\delta N = 1 - f(eU)$ follows the Fermi-Dirac distribution
function, one has the standard relation
\begin{equation}
C(U, T) = \frac{e^2}{k_B T} \delta N (1 - \delta N) .
\label{parabolic}
\end{equation}
Here we have assumed that $k_B T, e|U| \ll e^2/C_B$, such
that all higher energy charge configurations can be neglected.
Thus, up to a prefactor of $1/T$, $C(U, T)$ displays a
universal parabolic dependence on $\delta N$. The onset of
quantum fluctuations is reflected in the breakdown of the
above parabolic form.

As seen in Fig.~\ref{fig:fig9}, for $T \gg T_K$ the
capacitance line shape approaches the parabolic form of
Eq.~(\ref{parabolic}), in accordance with predominant thermal
fluctuations. By contrast, for $T < T_K$ there are strong
deviations from the parabolic form of Eq.~(\ref{parabolic}).
Specifically, the capacitance line shape is much narrower then
that of Eq.~(\ref{parabolic}), with concave shoulders that
saturate with decreasing temperature. This is most clearly
seen in the inset of Fig.~\ref{fig:fig9}, where the reduced
capacitance is plotted versus $\delta N$.

\section{Discussion}
\label{sec:discussion}

In this paper, we have extended the NCA to the Kondo-spin
Hamiltonian with arbitrary spin-exchange and
potential-scattering couplings, and applied it to the charge
fluctuations in a single-electron box at the Coulomb blockade.
The KNCA correctly describes the non-Fermi-liquid fixed point of
the multi-channel Kondo effect both qualitatively and quantitatively.
It reproduces the exact non-Fermi-liquid exponents and
logarithms of the
multi-channel Kondo effect, and gives surprisingly accurate
results for the temperature and field dependence of the magnetic
susceptibility in the isotropic two-channel case (see
Fig.~\ref{fig:fig5}). Hence the KNCA offers a reliable
approach to study the delicate interplay between the temperature
$T$, the local magnetic field $H$, and the Kondo temperature $T_K$
in the non-Fermi-liquid regime of the two-channel Kondo effect.

At the same time, the KNCA reveals several shortcomings of the
NCA that are not apparent from the more common applications of
this approach to the multi-channel Anderson and Coqblin-Schrieffer
Hamiltonians:

{\em 1. Kondo temperature} ---
For general antiferromagnetic couplings, the NCA class of
diagrams is insufficient for producing the correct
exponential dependence of $T_K$ on the inverse coupling
constants. In fact, for an isotropic spin-exchange interaction
$J$, the correct exponential dependence of $T_K$ on $1/J$
is recovered only if a matching potential-scattering term
of magnitude $J_0=J$ is included. Interestingly, this is
precisely the case corresponding to the Anderson and
Coqblin-Schrieffer Hamiltonians. That $J_0$ enters the
exponential dependence of $T_K$ is an artifact of the
KNCA, as is the fact that $J_0 = J$ is required in order
to recover the correct exponential dependence of $T_K$ on $1/J$.

{\em 2. Particle-hole symmetry} ---
The NCA inherently breaks particle-hole symmetry. Starting
from a particle-hole symmetric Hamiltonian, the KNCA violates
particle-hole symmetry, which is most clearly manifest in
the asymmetric energy dependence of the conduction-electron
$T$-matrix. We note that a similar asymmetry in the shape of
the Kondo resonance
was also found for the multi-channel Anderson model, where
it was attributed to the lack of an underlying
particle-hole symmetry.~\cite{CR93} From the KNCA we
conclude that such an asymmetry in the Kondo resonance
is an inherent feature of the NCA, independent of whether the
underlying Hamiltonian is particle-hole symmetric or not.

{\em 3. Ferromagnetic Kondo effect} --- One regime
obviously not accessible starting from an Anderson Hamiltonian
is that of ferromagnetic spin-exchange couplings. For
ferromagnetic couplings, the KNCA produces a spurious
Kondo-type effect, which extends to pure potential
scattering if $J_0 > 0$.

As discussed in detail in Appendix~\ref{app:appB}, all the above
deficiencies of the KNCA are a consequence of the omission of
particle-particle diagrams, which must be treated on equal
footing for the spin-$\frac{1}{2}$ Kondo model. To this
end, an extended scheme was devised in Appendix~\ref{app:appC},
based on a complete summation of all parquet diagrams.
Preliminary studies of the extended scheme seem to
suggest the amendment of the above NCA flaws. We further
expect it to remedy the well-known NCA failure to describe
the Fermi-liquid ground state of the single-channel Kondo
effect, much in the same way as the conserving $T$-matrix
approach does for the Anderson model.~\cite{CTMA} Although
we anticipate an intimate relation between our extended
scheme and the conserving $T$-matrix approach of Kroha
{\em et al.},~\cite{CTMA} we are unable to establish
any formal linkage between the two approaches.

Applying the KNCA to the charge fluctuations in a single-electron
box at the Coulomb blockade, two approaches have been proposed
to quantify the crossover from thermal fluctuations to quantum
fluctuations in the case of a narrow point contact: (i) By
plotting the reduced capacitance
$C(U, T)/C(0, T)$ versus appropriate scaling combinations of
the electrostatic potential $U$ and the temperature $T$; and
(ii) By plotting the reduced capacitance versus the average
number of excess electrons in the quantum box, $\delta N$. In
the latter case, pure thermal fluctuations are characterized
by a universal parabolic form, which breaks down as soon as
the temperature is lowered down to $T_K$. In the former
case, different scalings of $U$ with $T$ characterize the
high-temperature (thermal fluctuations) and the low-temperature
(quantum fluctuations) regimes. As seen in Fig.~\ref{fig:fig7},
for $e^2/C_B \gg T \gg T_K$ the reduced capacitance approaches
a universal function of $x_1 = eU/k_B T$, corresponding to the
derivative of the Fermi-Dirac distribution function. By
contrast, for $T_K \gg T$ there is an approximate scaling
with $x_2 = eU/k_B \sqrt{T_K T}$, which differs markedly from
both thermal fluctuations and ordinary lifetime broadening.

This unusual scaling of $U$ with $\sqrt{T}$ for $T \ll T_K$
is a distinct characteristic of the non-Fermi-liquid regime
of the two-channel Kondo effect. It directly probes the anomalous
one-half scaling dimension of an applied magnetic field
near the intermediate-coupling fixed point of the two-channel
model. Although the reduced capacitance does not show
exact scaling with $x_2$, the residual temperature dependence
is logarithmic in nature, and thus sufficiently weak to
render a scaling plot of $C(U, T)/ C(0, T)$ versus $U/\sqrt{T}$
a sharp experimental diagnostic for the observation of the
two-channel Kondo scenario. For
example, by plotting the reduced capacitance versus $U/\sqrt{T}$
for two temperatures differing by a factor of two or more,
one should be able to distinguish whether the relevant scaling
variable is $U/\sqrt{T}$ or not. The sole requirement is that
both temperatures be lower than $T_K$, which itself may be
extracted from the shape of the scaling curve. Moreover, the
proposed scaling relation should equally apply to the case of
nearly perfect transmission, which likewise corresponds to the
two-channel Kondo effect.~\cite{Matveev95}

A few words are in order at this point about the experimental
feasibility of reaching the low-temperature, non-Fermi-liquid
regime of the two-channel Kondo effect. There are two main
limiting factors for observing a fully developed
two-channel Kondo effect: (i) The charging energy, $e^2/C_B$,
must be sufficiently large in order for a measurable Kondo
temperature to emerge; (ii) The mean level spacing in the quantum
box must be sufficiently small as not to cut off the two-channel
Kondo effect. These two conditions set opposite limitations
on the size of the quantum box, which cannot be too large
nor too small. As recently pointed out by Zar\'and
{\em et al.},~\cite{ZZW00} it is practically impossible to
meet both conditions in present semiconducting devices, but
these conditions can be fulfilled in metallic quantum boxes.
The main complication with metallic devices has to do with
the construction of atomic-size junctions, which are required
for obtaining single-mode junctions. The latter are needed
since the relevant energy scale for the onset of the
two-channel Kondo effect decreases exponentially with the
number of tunneling modes.~\cite{ZZW00}

Finally, we wish to mention several potential applications of
the KNCA, beyond the present study of the charge fluctuations
in a quantum box. The most natural, perhaps, is the application
of the KNCA to the nonequilibrium scattering off nonmagnetic
two-level tunneling systems.~\cite{RB92,RLvDB94,von_Delft_et_al}
Previous efforts in this direction have focused on the
two-channel Anderson model,~\cite{HKH94} which has no direct
microscopic justification for TLS. The KNCA allows one to study
the actual anisotropic Kondo model appropriate for this case.
It would be particularly interesting to see whether there are any
quantitative changes to the scaling curves for the differential
conductance with $eV/k_BT$ upon going from the two-channel
Anderson model to the anisotropic two-channel Kondo model.

Another interesting possibility is the extension of the
present work to the calculation of the nonequilibrium
current through a single-electron transistor near the
degeneracy point. In this case, the single-electron box
is connected to two separate leads, in between which a
voltage bias is applied. The linear response of such a
system was computed for strong tunneling by Furusaki
and Matveev,~\cite{FM95} under the stringent assumption
that each lead is coupled to different conduction-electron
modes within the box (i.e., no coherent propagation between the
two leads~\cite{AN90}). This assumption, which corresponds
to a four-channel Kondo Hamiltonian, has far-reaching
implications for the zero-temperature conductance, which vanishes
at the degeneracy point for any asymmetry in the couplings
to the left and right leads.~\cite{FM95} In the general
case we expect, though, some overlap between the modes
coupled to the left and right leads, which should restore
the two-channel Kondo picture for tunneling at sufficiently
low temperatures. Whether this results in a square-root
voltage and temperature dependence of the differential
conductance as in scattering off TLS remains to be seen.

Lastly, one can exploit the KNCA to study the STM spectra
around an isolated Kondo impurity placed on a $d$-wave
superconductor. Such a setting may be realized in recent
STM experiments on Zn impurities in
Bi$_2$Sr$_2$CaCu$_2$O$_{8+\delta}$.~\cite{Zn_experiments}
While large-$N$ mean-field theories for the STM spectra around
an isolated magnetic adatom are expected to capture the main
qualitative features,~\cite{PSV00,ZT00} these approaches
are subject to various limitations, and are quantitatively
inaccurate. The KNCA, while not free of shortcomings, should
provide a more accurate framework for describing the STM
spectra.

\section*{Acknowledgements}
A.S. is grateful to Frithjof Anders, Daniel Cox, and
Matthias Hettler, for many useful discussions. E.L. and
A.S. were supported in part by the Centers of Excellence
Program of the Israel science foundation, founded by The
Israel Academy of Science and Humanities.

\appendix
\section{Details of the KNCA}
\label{app:appA}

\subsection{Derivation of the magnetic NCA equations}
In the presence of a nonzero magnetic field, spin up and spin down
are no longer equivalent. This complicates the structure of the
KNCA equations, which involve in addition to the $A_{\mu}$
coefficients of Eqs.~(\ref{A_0_def})--(\ref{A_i_def}) also the
following combinations:
\begin{eqnarray}
\Gamma_{0}=(J_0+J_z)/4,\;\;\;\;
\Gamma_{\perp}=(J_x+J_y)/4,\\
\Lambda_{0}=(J_0-J_z)/4,\;\;\;\;
\Lambda_{\perp}=(J_x-J_y)/4.
\end{eqnarray}

\begin{figure}
\centerline{
\vbox{ \epsfxsize=70mm \epsfbox {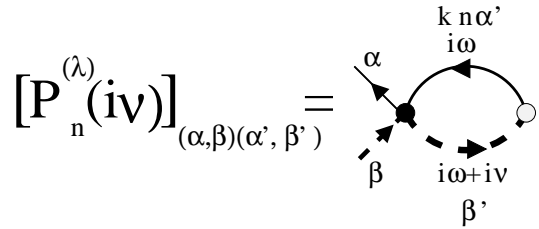}}
}\vspace{10pt}
\caption{Diagrammatic representation of $P^{(\lambda)}(i\nu)$,
         which serves as the building block of the NCA self-energy
	 diagram of Fig.~\ref{fig:fig3}. $P^{(\lambda)}(i\nu)$
	 consists of a particle-hole bubble and a bare vertex
	 attached at one end. Here $\alpha, \beta, \alpha'$,
	 and $\beta'$ are spin labels, $n$ is a conduction-electron
	 channel index, and there is an implicit summation over
	 the conduction-electron momentum index $k$. Dashed lines
         represent fully dressed $f$ Green functions.}
\label{fig:app1}
\end{figure}

Adopting the $4\!\times\!4$ matrix convention of
Eqs.~(\ref{V_abgd_1})--(\ref{V_abgd_2}), Fig.~\ref{fig:app1}
shows the building block of the NCA self-energy diagrams of
Fig.~\ref{fig:fig3}, which we denote by
$[P_{n}^{(\lambda)}(i\nu)]_{(\alpha,\beta)(\alpha',\beta')}$.
It consists of a particle-hole bubble and a bare vertex attached at
one end. Upon projection,
$P_{n}(z) = \lim_{\lambda \to -\infty} P_n^{(\lambda)}(z - \lambda)$
takes the form
\begin{equation}
P_{n}(z) = \left [
         \begin{array}{cccc}
                \Gamma_{0}\Pi_{\uparrow}(z) & 0 & 0 &
                        \Gamma_{\perp}\Pi_{\downarrow}(z) \\
                0 & \Lambda_{0}\Pi_{\downarrow}(z) &
                        \Lambda_{\perp}\Pi_{\uparrow}(z) & 0 \\
                0 & \Lambda_{\perp}\Pi_{\downarrow}(z) &
                        \Lambda_{0}\Pi_{\uparrow}(z) & 0 \\
                \Gamma_{\perp}\Pi_{\uparrow}(z) & 0 & 0 &
                        \Gamma_{0}\Pi_{\downarrow}(z)
         \end{array} \right ],
\label{P^lambda}
\end{equation}
where
\begin{equation}
\Pi_{\sigma}(z) = - \sum_k f(\epsilon_k) G_{\sigma}(z + \epsilon_k)
\label{Pi_bubble}
\end{equation}
is the projected bubble. Here we have written the matrix $P_{n}(z)$
with respect to the basis $(\uparrow, \uparrow) \leftrightarrow 1$,
$(\downarrow, \uparrow) \leftrightarrow 2$,
$(\uparrow, \downarrow) \leftrightarrow 3$, and
$(\downarrow, \downarrow) \leftrightarrow 4$, and made use of the
fact the pseudo-fermion Green function is diagonal in the spin
index. Note that $P$ does not depend on the channel index $n$,
which is omitted hereafter.

Using the above matrix convention, the ladder diagram of
Fig.~\ref{fig:fig3} is conveniently reduced to a geometric series
in the matrix $P$. The resulting expression for the projected
pseudo-fermion self-energy reads~\cite{comment_on_constant_part}
\begin{equation}
\Sigma_{\sigma}(z) = -M\!\sum_{\alpha = \uparrow, \downarrow} \sum_k
       f(-\epsilon_k)\!\left [ \frac{1}{1 - P(z-\epsilon_k)}V
       \right ]_{(\alpha\sigma)(\alpha\sigma)} ,
\label{Sigma_via_P}
\end{equation}
where $M$ is the number of conduction-electron channels, and $V$ is the
interaction matrix of Eq.~(\ref{V_abgd_2}). The final task in
deriving the KNCA equations is to invert the $1 - P$ matrix in
Eq.~(\ref{Sigma_via_P}). For a zero magnetic field, when
$\Pi_{\uparrow}(z)$ and $\Pi_{\downarrow}(z)$ are equal, one readily
obtains Eqs.~(\ref{sigma_H=0})--(\ref{Pi_H=0}). For a nonzero magnetic
field, when $\Pi_{\uparrow}(z)$ and $\Pi_{\downarrow}(z)$ differ, a more
complicated expression is obtained. Explicitly, $\Sigma_{\sigma}(z)$
takes the form
\begin{equation}
\Sigma_{\sigma}(z) = \frac{M}{2}
                   \sum_k f(-\epsilon_k)\Delta_{\sigma}(z - \epsilon_k) ,
\label{Sigma_general}
\end{equation}
where
\begin{eqnarray}
&& \Delta_{\sigma}(z) =
    \frac{-2\Gamma_{0} + 2(\Gamma_{0}^2 - \Gamma_{\perp}^2)
    \Pi_{-\sigma}(z)}{ [1 - \Gamma_{0} \Pi_{\uparrow}(z)]
                       [1 - \Gamma_{0} \Pi_{\downarrow}(z)]
                       - \Gamma_{\perp}^2 \Pi_{\uparrow}(z)
                       \Pi_{\downarrow}(z) } \nonumber\\
&& \;\;\;
+ \frac{-2\Lambda_{0} +
       2(\Lambda_{0}^2 - \Lambda_{\perp}^2)\Pi_{-\sigma}(z)}
       { [1 - \Lambda_{0} \Pi_{\uparrow}(z)]
         [1 - \Lambda_{0} \Pi_{\downarrow}(z)]
         - \Lambda_{\perp}^2 \Pi_{\uparrow}(z) \Pi_{\downarrow}(z) } .
\label{Delta_general}
\end{eqnarray}
Equations~(\ref{Pi_bubble}), (\ref{Sigma_general}), and
(\ref{Delta_general}) constitute the full set of KNCA equations
for a general nonzero magnetic field. For a zero magnetic field,
they properly reduce to Eqs.~(\ref{sigma_H=0})--(\ref{Pi_H=0}).

\subsection{Negative-frequency spectral functions}
Although all relevant physical information is contained in
principle in the ordinary pseudo-fermion spectral functions,
numerical considerations dictate the introduction of yet another
set of functions. To see this we note that the evaluation of
any physical observable necessarily involves the integral of
a pseudo-fermion spectral function multiplied by a Boltzmann
factor that diverges for large negative energies. Hence the
dominant contribution to such integrals comes from a regime
where the pseudo-fermion
spectral functions are exponentially small. Since it is numerically
impossible to maintain such an exponential accuracy when solving
the ordinary set of KNCA equations, an alternative approach is
required. This is achieved by absorbing the above-mentioned
Boltzmann factor into the definition of the negative-frequency
spectral functions,
\begin{equation}
a_{\sigma}(\epsilon) = - \frac{1}{\pi} e^{\beta(\epsilon_0 - \epsilon)}
        {\rm Im} \left \{G_{\sigma}(\epsilon + i\delta) \right\} ,
\label{Anegdef}
\end{equation}
and deriving an alternative set of equations directly for the
functions $a_{\sigma}(\epsilon)$. In the above equation,
$\beta = 1/k_B T$ is the inverse temperature, and $\epsilon_0$ is
an arbitrary numerical parameter, chosen to avoid exponentially
large numbers.~\cite{comment_on_e_0} The main physical quantities
of interest are conveniently expressed in terms of the
negative-frequency spectral functions. Specifically, the
impurity contribution to the partition function, Eq.~(\ref{Z_imp}),
is given by
\begin{equation}
Z_{imp} = e^{-\beta\epsilon_0}
          \int_{-\infty}^{\infty} \left [a_{\uparrow}(\epsilon)
          + a_{\downarrow}(\epsilon) \right ] d\epsilon,
\end{equation}
while the impurity magnetization is expressed as
\begin{equation}
M(H,T) = \frac{1}{2}\mu_Bg_J \frac{\int_{-\infty}^{\infty}
         \left [ a_{\uparrow}(\epsilon)
               - a_{\downarrow}(\epsilon) \right ] d\epsilon}
         {\int_{-\infty}^{\infty} \left [ a_{\uparrow}(\epsilon)
               + a_{\downarrow}(\epsilon) \right ] d\epsilon}.
\label{Magnetization}
\end{equation}

To derive an alternative set of equations for the negative-frequency
spectral functions, it is necessary to take the imaginary parts of
Eqs.~(\ref{Pi_bubble}), (\ref{Sigma_general}) and
(\ref{Delta_general}), and explicitly multiply them by the Boltzmann
factor $e^{\beta (\epsilon_0 - \epsilon)}$. Upon doing so, one
arrives at the following set of KNCA equations for the
negative-frequency functions:
\begin{equation}
a_{\sigma}(\epsilon) = \frac{M}{2} |G_{\sigma}(\epsilon)|^2 \sum_k
          f(\epsilon_k) d_{\sigma}(\epsilon - \epsilon_k)
\end{equation}
with
\begin{eqnarray}
d_{\sigma}(\epsilon) &=& \varphi_{\sigma}(\epsilon)
        \sum_k f(-\epsilon_k) a_{\sigma}(\epsilon + \epsilon_k)
\nonumber\\
&+& \psi_{\sigma}(\epsilon)
        \sum_k f(-\epsilon_k) a_{-\sigma}(\epsilon + \epsilon_k) .
\label{d_def}
\end{eqnarray}
Here $\varphi_{\sigma}(\epsilon)$ and $\psi_{\sigma}(\epsilon)$ are two
rather complicated expressions given by
\begin{eqnarray}
&& \varphi_{\sigma}(\epsilon) =
        \frac{ 2C^2(\Gamma_{0}, \Gamma_{\perp}, \Pi_{-\sigma}) }
        {\left | (1 - \Gamma_{0} \Pi_{\uparrow})
                                (1 - \Gamma_{0} \Pi_{\downarrow})
                                - \Gamma_{\perp}^2 \Pi_{\uparrow}
                                \Pi_{\downarrow} \right |^2 }
\label{phi_aux} \\
\nonumber\\
&&+\; ({\rm the \; same \; with \;}
       \Gamma_{0}, \Gamma_{\perp} {\rm \; and \;}
       \Lambda_{0},\Lambda_{\perp}\;{\rm interchanged}) \nonumber
\end{eqnarray}
and
\begin{eqnarray}
&&\psi_{\sigma}(\epsilon) =
             \frac{ 2C(\Gamma_{0}, \Gamma_{\perp}, \Pi_{-\sigma})
                     C(\Gamma_{0}, \Gamma_{\perp}, \Pi_{\sigma}) }
                  {\left | (1 - \Gamma_{0} \Pi_{\uparrow})
                     (1 - \Gamma_{0} \Pi_{\downarrow})
                      - \Gamma_{\perp}^2 \Pi_{\uparrow}
                           \Pi_{\downarrow} \right |^2 }
\label{psi_aux} \\
&&+\; 2(\Gamma_{0}^2\!-\!\Gamma_{\perp}^2)
        {\rm Re} \left \{ \frac{1}
        { (1\!-\!\Gamma_{0} \Pi_{\uparrow})
        (1\!-\!\Gamma_{0} \Pi_{\downarrow})
         - \Gamma_{\perp}^2 \Pi_{\uparrow} \Pi_{\downarrow} } \right \}
\nonumber\\
\nonumber\\
&&+\; ({\rm the \; same \; with \;}
       \Gamma_{0}, \Gamma_{\perp} {\rm \; and \;}
       \Lambda_{0},\Lambda_{\perp}\;{\rm interchanged}) ,\nonumber
\end{eqnarray}
where
\begin{equation}
C(x, y, z) = x + ( y^2 - x^2 ) {\rm Re} \{z\} .
\end{equation}
In Eqs.~(\ref{phi_aux})--(\ref{psi_aux}), we have omitted for
conciseness the arguments of $\Pi_{\uparrow}(\epsilon + i\delta)$
and $\Pi_{\downarrow}(\epsilon + i\delta)$.

\begin{figure}
\centerline{
\vbox{ \epsfxsize=80mm \epsfbox {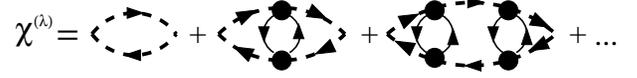}}
}\vspace{10pt}
\caption{NCA diagrams for the zero-field susceptibility of a Kondo
         impurity, as derived from the Kubo formula. Here filled
	 circles represent dressed vertices (see Fig.~\ref{fig:app5}),
	 while dashed lines stand for fully dressed $f$ Green
	 functions. In the Coqblin-Schrieffer limit, all higher
	 order terms vanish and only the bare bubble is left.
	 In contrast, all terms contribute to the susceptibility
	 of a generic Kondo Hamiltonian.}
\label{fig:app2}
\end{figure}

Note that the negative-frequency equations explicitly depend on
$G_{\sigma}(z)$ and $\Pi_{\sigma}(z)$, hence the negative-frequency
and ordinary KNCA equations must be solved simultaneously. This is
done by repeated numerical iterations until convergence is reached.
As is always the case with numerical solutions of NCA-type equations,
the key to high-precision numerics is in a well-designed grid of mesh
points that scatter mesh points more densely near the threshold and
peaks of the relevant functions. To this end, we have used a
combination of linear and logarithmic grids. As a critical test for
the precision of our numerical code, we have checked in all our runs
that the pseudo-fermion spectral functions fulfilled the spectral
sum rule to within one part in one thousand (i.e., $0.1$\%).

\subsection{Magnetic vertex correction}
As mentioned in the main text, we have calculated the magnetic
susceptibility $\chi(H, T)$ by numerically differentiating the
magnetization of Eq.~(\ref{Magnetization}) with respect to $H$.
An interesting observation has to do with the calculation of
the zero-field susceptibility, $\chi(T)$. For the Anderson and
Coqblin-Schrieffer models, one traditionally computes the
zero-field susceptibility from the Kubo formula, which reduces
in the NCA to a simple bubble diagram.~\cite{Bickers87}
For a Kondo model, however, there is a nontrivial magnetic
vertex correction that cannot be neglected. Fig.~\ref{fig:app2}
depicts the NCA susceptibility diagrams for a generic Kondo model.
In addition to the simple bubble, there is a magnetic vertex correction
in the form of a ladder diagram. Each rung of the ladder consists
of two dressed vertices (displayed in Fig.~\ref{fig:app5}),
connected by a conduction-electron
bubble. Since the rungs are frequency dependent, a direct evaluation
of the entire series requires the solution of an integral equation,
which is far more complicated than numerical differentiation of the
magnetization curve.

\begin{figure}
\centerline{
\vbox{\epsfxsize=85mm \epsfbox {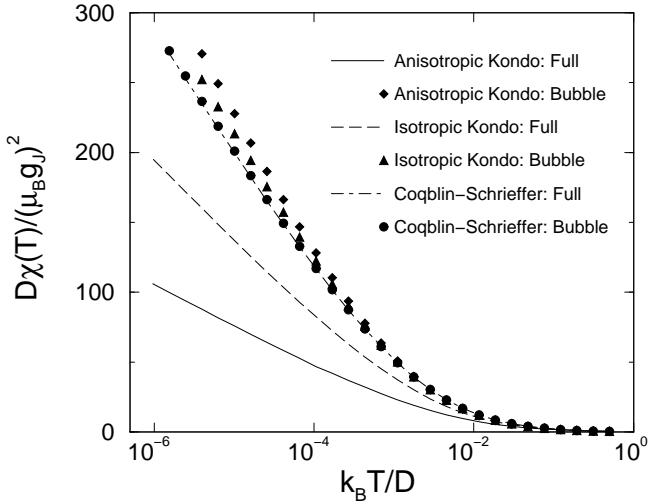}}
}\vspace{10pt}
\caption{Comparison between the simple magnetization bubble
         and the full zero-field susceptibility, obtained by
	 differentiating $M(H, T)$ with respect to $H$.
         All curves refer to the two-channel case, with
	 $\rho_0 A_0$ held fixed at $0.2$. While the two
	 procedures give identical results in the Coqblin-Schrieffer
	 limit (we use $\rho_0 J_0 = \rho_0 J = 0.2$), substantial
	 deviations are found for the isotropic and anisotropic
	 Kondo models (we use $\rho_0 J = 0.2667$ for the
         isotropic Kondo model, and $J_z = 0$ and
	 $\rho_0 J_{\perp} = 0.4$ for the anisotropic Kondo Model).
	 In the latter two cases, the effect of the magnetic vertex
	 correction is to significantly reduce the slope of the
	 log-$T$ diverging term in $\chi(T)$.}
\label{fig:app3}
\end{figure}

While in the Coqblin-Schrieffer limit, $J_0 = J_x = J_y = J_z$, it
possible to show that only the simple bubble survives,~\cite{comment_on_CS}
for a generic Kondo model there are non-negligible contributions coming
from the magnetic vertex correction. For the two-channel case, this is
demonstrated in Fig.~\ref{fig:app3}, where the results of the
bare bubble are compared with the full zero-field susceptibility
obtained by differentiating $M(H, T)$. As required, the two
procedures are indistinguishable in the Coqblin-Schrieffer limit,
establishing the numerical accuracy of our differentiation routine.
A different picture is recovered, however, for the isotropic and
anisotropic Kondo models. In both cases the slope of the log-$T$
diverging term in $\chi(T)$ is significantly reduced upon going
from the simple bubble diagram to the full susceptibility. This
effect is particularly pronounced for the anisotropic Hamiltonian
of Eq.~(\ref{H_matveev}), where nearly a three-fold reduction is
found in the slope of the log-$T$ component of $\chi(T)$. Thus,
the magnetic vertex correction contains essential contributions to
the low-temperature susceptibility of the two-channel Kondo model.

\begin{figure}
\centerline{
\vbox{\epsfxsize=80mm \epsfbox {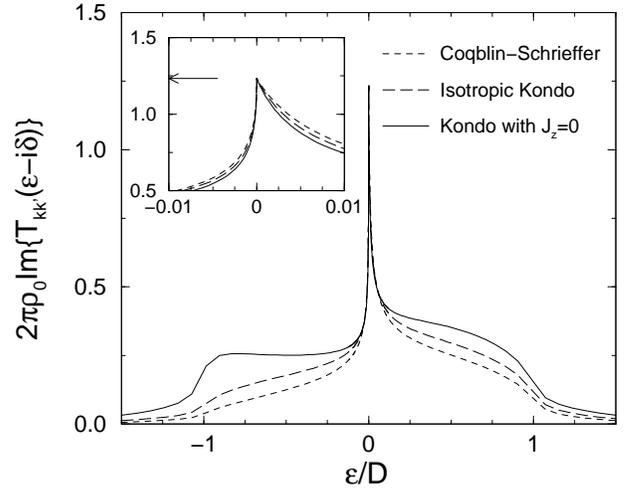}}
}\vspace{10pt}
\caption{Imaginary part of the conduction-electron $T$-matrix, as
         obtained from the KNCA. Here, the solid line corresponds
	 to the anisotropic two-channel Kondo model with $J_z = 0$
	 and $\rho_0 J_{\perp} = 0.4$
	 ($k_B T_K/D = 3.85\!\times\!10^{-3}$); the long-dashed
	 line is for the isotropic two-channel Kondo model with
	 $\rho_0 J = 0.2667$ ($k_B T_K/D = 2\!\times\!10^{-3}$);
	 and the short-dashed line is for the two-channel
	 Coqblin-Schrieffer model with $\rho_0 J_0 = \rho_0 J = 0.2$
	 ($k_B T_K/D = 1.35\!\times\!10^{-3}$). The temperature
	 in all curves is equal to $T/T_K= 0.01$. In accordance
	 with the analytic treatment of the NCA equations at zero
         temperature,~\protect\cite{CR93} the $T$-matrix has a
	 cusp at zero energy, with a strong particle-hole asymmetry.
	 All curves fall within $1.5\%$ from the predicted
	 zero-temperature NCA value of
	 $2\pi\rho_0 {\rm Im}\{ T_{kk'}(0-i\delta) \} = \pi^2/8$
         (marked by an arrow in the inset).}
\label{fig:app4}
\end{figure}

\subsection{Conduction-electron $T$-matrix}
Another quantity of interest in the theory of dilute magnetic alloys
is the conduction-electron $T$-matrix, defined by the expansion
\begin{equation}
G_{kk'\!,\sigma}(z) = G^{(0)}_{k\sigma}(z) \delta_{k,k'} +
      G^{(0)}_{k\sigma}(z) T_{kk'\!,\sigma}(z) G^{(0)}_{k'\sigma}(z) .
\label{T_matrix}
\end{equation}
Here $G_{kk'\!,\sigma}(z)$ is the Fourier representation of the dressed
conduction-electron Green function
\begin{equation}
G_{kk'\!,\sigma}(\tau) = - \langle T_{\tau} c_{k n \sigma}(\tau)
                     c^{\dagger}_{k' n \sigma}(0) \rangle ,
\end{equation}
and $G^{(0)}_{k\sigma}(z) = 1/(z - \epsilon_k)$ is the bare
conduction-electron Green function. For a nondegenerate Anderson
impurity, $T_{kk'\!,\sigma}(z)$ is equal to $V^2 G_{imp,\sigma}(z)$,
where $V$ is the hybridization matrix element (assumed here to be $k$
independent), and $G_{imp,\sigma}(z)$ is the dressed impurity Green
function. Hence the calculation of the $T$-matrix is reduced
to that of the dressed impurity Green function. For a
Kondo impurity there is no analogous relation, and $T_{kk'\!,\sigma}(z)$
needs to be constructed directly from Eq.~(\ref{T_matrix}).

Within the slave-fermion representation, the $T$-matrix is
related to the {\em unprojected} conduction-electron self-energy
through
\begin{equation}
T_{kk'\!,\sigma}(z) = \frac{1}{Z_{imp}} 
        \lim_{\lambda \rightarrow -\infty}
        e^{-\beta\lambda} \Sigma_{kk'\!,\sigma}^{(\lambda)}(z) .
\end{equation}
Using the NCA class of diagrams for the conduction-electron
self-energy, and in the absence of an applied magnetic field,
one obtains
\begin{equation}
T_{kk'}(z) = \frac{e^{-\beta \epsilon_0}}{2 Z_{imp}}
         \int_{-\infty}^{\infty} \left [ d(\epsilon) G(z + \epsilon)
         - a(\epsilon) \Delta(\epsilon - z) \right] d\epsilon ,
\label{T_matrix_NCA}
\end{equation}
where $G(z)$ is the pseudo-fermion Green function, $a(\epsilon)$
is the negative-frequency spectral function of Eq~(\ref{Anegdef}),
$\Delta(z)$ is defined in Eq.~(\ref{Delta_def}), and $d(\epsilon)$
is given by Eq.~(\ref{d_def}). Note that all spin indices have
been dropped in Eq.~(\ref{T_matrix_NCA}) due to the equivalence
of the two spin orientations for a zero magnetic field. Also
notice that $T_{kk'}(z)$ has no explicit dependence on $k$
and $k'$, as is characteristic of purely local scattering.

Figure~\ref{fig:app4} shows the imaginary part of the KNCA
conduction-electron $T$-matrix, for each of the Coqblin-Schrieffer,
isotropic Kondo, and anisotropic Kondo models. The coupling
constants were adjusted in each model as to give $\rho_0 A_0 = 0.2$,
while the temperature $T/T_K = 0.01$ is sufficiently low in order for
fully developed Kondo peaks to be seen. Two points are noteworthy.
First, there is a clear asymmetry in the energy dependence
of ${\rm Im} \{ T_{kk'}(\epsilon + i\delta)\}$, which for the
isotropic and anisotropic Kondo models is in violation
of the underlying particle-hole symmetry. This asymmetry persists
from high energies, $|\epsilon| \sim D$, down to at low energies,
$|\epsilon| \alt T_K$. As discussed in Appendix~\ref{app:appB},
it stems from the omission of particle-particle bubbles within
the NCA. Thus, the asymmetric line shape of the Kondo resonance is an
inherent feature of the NCA class of diagrams, and is not due to
the lack of an underlying particle-hole symmetry, as previously
asserted for the multi-channel Anderson model.~\cite{CR93}

The second point to notice is that all curves approach the predicted
zero-temperature NCA value of
$2\pi\rho_0 {\rm Im}\{ T_{kk'}(0-i\delta) \} = \pi^2/8$.~\cite{CR93}
Since this value is determined by the NCA threshold exponents, which
are unchanged in going from the two-channel Anderson to the isotropic
and anisotropic Kondo models, then the same zero-temperature value
applies to all models.

\section{Comparison with poor-man's scaling}
\label{app:appB}

\begin{figure}
\centerline{
\vbox{ \epsfxsize=70mm \epsfbox {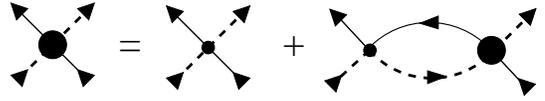}}
}\vspace{10pt}
\caption{Renormalization of the bare interaction within the KNCA.
         In the diagrams of Fig~\ref{fig:fig3}, the bare vertex is
	 dressed by a ladder of particle-hole bubbles, each composed
	 of one (bare) conduction-electron line and one (fully
	 dressed) pseudo-fermion line.}
\label{fig:app5}
\end{figure}

\begin{figure}
\centerline{
\vbox{ \epsfxsize=70mm \epsfbox {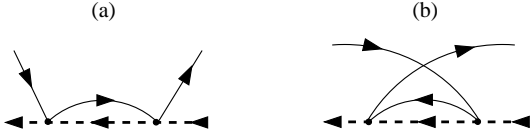}}
}\vspace{10pt}
\caption{Renormalization of the interaction within poor-man's
         scaling. We distinguish between (a) processes involving the
         excitation of a high-energy electron (particle-hole bubble)
         and (b) processes involving the excitation of a high-energy
         hole (particle-particle bubble). The KNCA retains only the
         particle-hole bubble of (a), albeit with self-consistent
         dressing of the pseudo-fermion Green functions.}
\label{fig:app6}
\end{figure}

Considerable insight into the origin of the NCA setbacks
can be gained by comparing the KNCA with Anderson's poor-man's
scaling.~\cite{Anderson70} Specifically, the KNCA set of
diagrams can be viewed as a ladder renormalization of
the bare interaction, as illustrated in Fig.~\ref{fig:app5}.
The basic renormalization of the bare vertex is due to
particle-hole bubbles. This should be contrasted with poor-man's
scaling, where two distinct types of renormalizations are
found: particle-hole and particle-particle bubbles (see
Fig.~\ref{fig:app6}). Thus, it appears that the KNCA corresponds
to retaining only the particle-hole processes within poor-man's
scaling (although additional higher-order terms do enter the KNCA
through the self-consistent dressing of the pseudo-fermion Green
functions).

To substantiate this correspondence between the KNCA and
poor-man's scaling, we have rewritten the well-known
renormalization-group (RG) equations for the dimensionless
coupling constants $\tilde{J}_{\mu} = \rho_0 J_{\mu}$
($\mu = 0, x, y, z$), by separating the
contributions of the two diagrams of Fig.~\ref{fig:app6}. Labeling
the contributions of the particle-hole and particle-particle diagrams
by $ph$ and $pp$, respectively, the corresponding equations read
\begin{eqnarray}
\left. \frac{d{\tilde J}_i}{dl}\right|_{ph} &=&
             \frac{1}{2} \left [ {\tilde J}_j {\tilde J}_k +
             {\tilde J}_0 {\tilde J}_i \right ] ,
\label{p-h_J} \\
\left. \frac{d{\tilde J}_0}{dl}\right|_{ph} &=&
            \frac{1}{4} \sum_{\mu = 0, x, y, z}{\tilde J}_{\mu}^2 ,
\label{p-h_V}
\end{eqnarray}
and
\begin{eqnarray}
\left. \frac{d{\tilde J}_i}{dl}\right|_{pp} &=&
             \frac{1}{2} \left [ {\tilde J}_j {\tilde J}_k -
             {\tilde J}_0 {\tilde J}_i \right ] ,
\label{p-p_J} \\
\left. \frac{d{\tilde J}_0}{dl}\right|_{pp} &=&
            -\frac{1}{4} \sum_{\mu = 0, x, y, z}{\tilde J}_{\mu}^2 ,
\label{p-p_V}
\end{eqnarray}
where $l$ is equal to $\ln (D_0/D)$, and $i, j$, and $k$ represent
a cyclic permutation of $x, y$, and $z$. Here $D$ is the running
bandwidth, and $D_0$ is the bare bandwidth. By combining the two
sets of equations listed above, one recovers the
standard RG equations:~\cite{Anderson70}
\begin{eqnarray}
\frac{d{\tilde J}_i}{dl} &=& {\tilde J}_j {\tilde J}_k ,
\label{full_RG_J}\\
\frac{d{\tilde J}_0}{dl} &=& 0 .
\label{full_RG_V}
\end{eqnarray}

If we artificially ignore the particle-particle contributions of
Eqs.~(\ref{p-p_J})--(\ref{p-p_V}) and settle with
Eqs.~(\ref{p-h_J})--(\ref{p-h_V}) for $d{\tilde J}_{\mu}/dl$,
it is easy to verify that
\begin{eqnarray}
{\tilde A}_0 &=& \frac{1}{4} \left ( {\tilde J}_0 + {\tilde J}_x +
                 {\tilde J}_y + {\tilde J}_z \right) ,\\
{\tilde A}_i &=& \frac{1}{4} \left ({\tilde J}_0 +  2{\tilde J}_i -
                 {\tilde J}_x - {\tilde J}_y - {\tilde J}_z \right)
\end{eqnarray}
($i = x, y, z$) obey the single differential equation
\begin{equation}
\left. \frac{d{\tilde A}_{\mu}}{dl} \right|_{ph} = {\tilde A}_{\mu}^2 .
\label{d_A_mu}
\end{equation}
Equation~(\ref{d_A_mu}) has the exact same form as the standard RG
equation for the dimensionless exchange coupling in the isotropic
Kondo model.~\cite{Anderson70} As is well known, its outcome depends
on the sign of the bare coupling constant. If ${\tilde A}_{\mu}$
is initially positive (``antiferromagnetic'' coupling) then this
coupling flows towards strong coupling; otherwise (``ferromagnetic''
coupling) it flows to weak coupling. In the former case, a
nonperturbative energy scale
$T_{\mu} \sim D_0 \exp \left [ -1/ \rho_0 A_{\mu} \right ]$ emerges,
below which ${\tilde A}_{\mu}$ of Eq.~(\ref{d_A_mu}) becomes of
order unity. If several ${\tilde A}_{\mu}$ simultaneously flow
towards strong coupling, then the largest of the $T_{\mu}$'s
will emerge as the actual low-energy scale.

From this analysis it is clear that Eqs.~(\ref{p-h_J})--(\ref{p-h_V})
flow towards strong coupling whenever at least one of the $A_{\mu}$'s
of Eqs.~(\ref{A_0_def})--(\ref{A_i_def}) is positive. The underlying
nonperturbative energy scale that emerges then is given by
\begin{equation}
T_K^{(ph)} \sim D_0 \exp(-\frac{1}{\rho_0  A_{\rm max} }),
\label{T_A_max}
\end{equation}
where $A_{\rm max}$ is the largest of the positive $A_{\mu}$'s.
Note that for antiferromagnetic couplings with $J_{\perp} > 0$,
$A_{\rm max}$ is always given by $A_0$.

An identical picture is recovered within the KNCA equations for a
zero magnetic field, Eqs.~(\ref{sigma_H=0})--(\ref{Pi_H=0}). Here
the flow towards strong coupling is manifest in the divergence of
$\Delta(\epsilon + i\delta)$ at the threshold energy as
$T \to 0$,~\cite{comment_on_w_c} which happens whenever at least
one of the $A_{\mu}$'s is positive. If more than one $A_{\mu}$
coefficient is positive then only that
$A_{\mu}/[1 - A_{\mu}\Pi(\epsilon + i\delta)]$ component of
$\Delta(\epsilon + i\delta)$ corresponding to $A_{\rm max}$ diverges,
while all other components remain finite.~\cite{comment_on_A_m}
The nonperturbative energy scale underlying
the NCA is then precisely given by Eq.~(\ref{T_A_max}),
establishing the intimate connection between the KNCA
and the restricted version of poor-man's scaling where only
particle-hole processes are retained. (Obviously, the two
approaches are not entirely equivalent, as the KNCA also
includes self-consistent dressing of the pseudo-fermion
Green functions.)

\begin{figure}
\centerline{
\vbox{ \epsfxsize=55mm \epsfbox {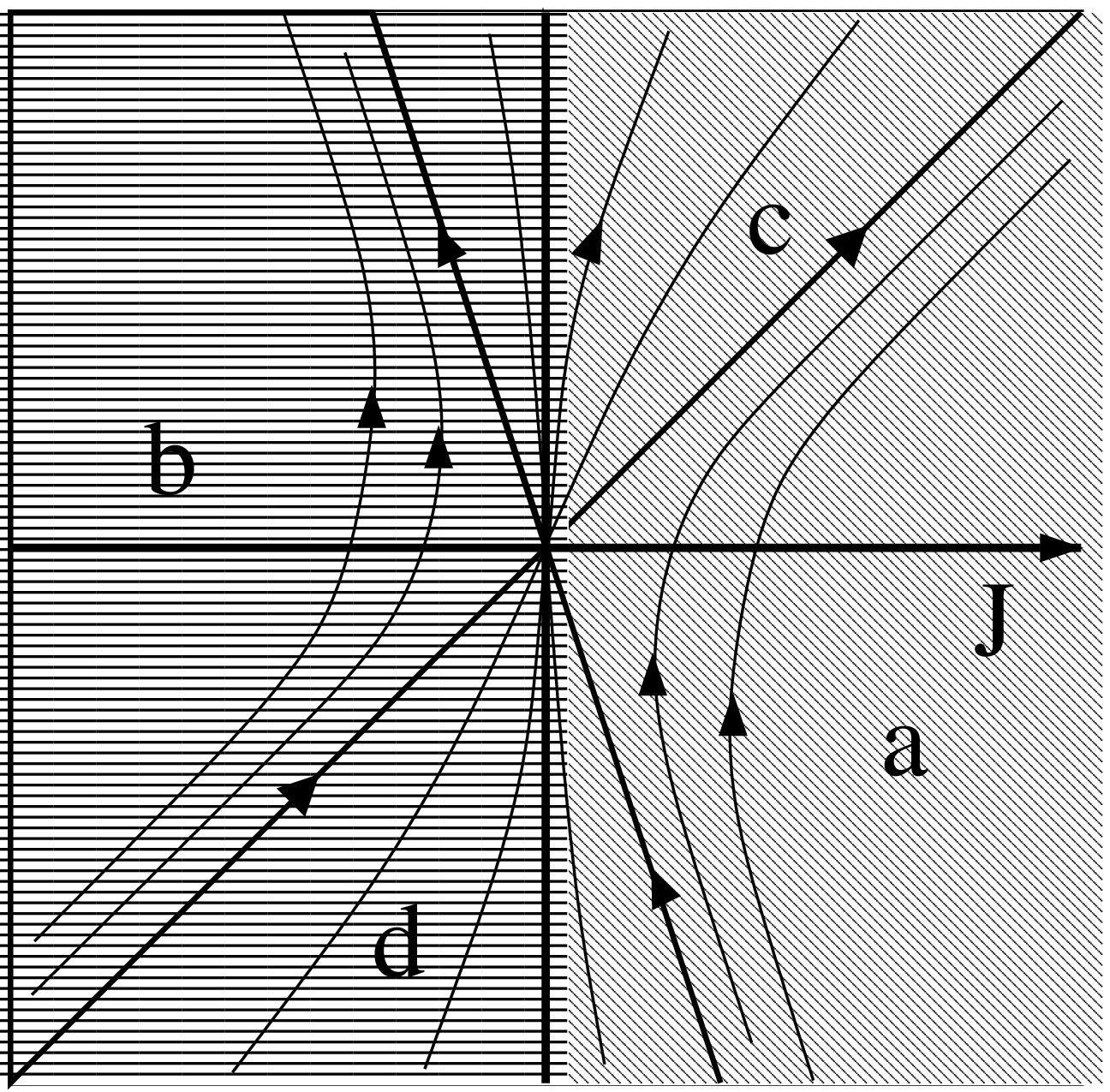}}
}\vspace{10pt}
\caption{RG flow trajectories corresponding to
         Eqs.~(\ref{p-h_J})--(\ref{p-h_V}) with an isotropic
         spin-exchange coupling $J$. There are four different
	 RG regimes: (a) $-3J < J_0 \le J$, where only
	 $\tilde{A}_0$ is relevant; (b) $J < J_0 \le -3J$,
	 where only $\tilde{A}_0$ is irrelevant; (c)
	 $-3J, J < J_0$, where all the $\tilde{A}_{\mu}$'s
         are relevant; and (d) $J_0 \le -3J, J$, where all
	 the $\tilde{A}_{\mu}$'s are irrelevant. The flow
	 in region (a) and in the $J > 0$ part of region
	 (c) is towards $J_0, J = \infty$ and $J_0/J = 1$;
	 in region (b) and in the $J < 0$ part of region
         (c) the flow is towards $J_0, -J = \infty$ and
	 $J_0/J = -3$; the flow along the $J = 0$ line in
	 region (c) is towards $J_0 = \infty$ and $J = 0$;
	 and in region (d) the flow is towards $J_0 = J = 0$.}
\label{fig:app7}
\end{figure}

We can now exploit this basic relation between the KNCA and
Eqs.~(\ref{p-h_J})--(\ref{p-h_V}) to clarify the setbacks
of the KNCA. For simplicity, let us focus on the isotropic
case, $J_x = J_y = J_z = J$. From Eq.~(\ref{p-h_V}) one
readily sees that, starting from $J_0 = 0$, a nonzero
potential scattering is immediately generated. This is
to be contrasted with the full set of RG equations, in
which potential scattering is exactly marginal. Thus,
starting from a particle-hole symmetric Hamiltonian, the
omission of the particle-particle contributions necessarily
breaks particle-hole symmetry, leading, for example, to
the asymmetric energy dependence of the conduction-electron
$T$-matrix within the KNCA (see Fig~\ref{fig:app4}).

More significantly, the interplay between the spin-exchange couplings
and the ever-increasing potential scattering substantially alters the
RG flow diagram. In Fig.~\ref{fig:app7}, the RG flow trajectories
corresponding to Eqs.~(\ref{p-h_J})--(\ref{p-h_V}) are plotted
in the $J-J_0$ plane. While the full set of RG
equations, Eqs.~(\ref{full_RG_J})--(\ref{full_RG_V}), support
only horizontal trajectories, non of the trajectories of
Eqs.~(\ref{p-h_J})--(\ref{p-h_V}) are horizontal. Overall
there are four different regimes: (a) $-3J < J_0 \le J$, where
only $\tilde{A}_0$ is relevant; (b) $J < J_0 \le -3J$, where
only $\tilde{A}_0$ is irrelevant; (c) $-3J, J < J_0$, where
all the $\tilde{A}_{\mu}$'s are relevant; and (d) $J_0 \le -3J, J$,
where all the $\tilde{A}_{\mu}$'s are irrelevant. Thus, with
the exception of region (d), Eqs.~(\ref{p-h_J})--(\ref{p-h_V})
flow towards some kind of strong coupling, including in the case of
ferromagnetic coupling. The emergence of a ferromagnetic Kondo-type
effect within the KNCA is therefore an artifact of the
omission of particle-particle diagrams.

Finally, if one naively extends Eqs.~(\ref{p-h_J})--(\ref{p-h_V})
to the strong-coupling regime, then four different fixed points
are recovered:
(i) $J_0, J = \infty$ and $J_0/J = 1$ [approached
    from region (a) and from the $J > 0$ part of region (c)];
(ii) $J_0, -J = \infty$ and $J_0/J = -3$ [approached
    from region (b) and from the $J < 0$ part of region (c)];
(iii) $J_0 = \infty$ and $J = 0$ [approached
    from the $J = 0$ line in region (c)];
and (iv) $J_0 = J = 0$ [approached from region (d)].
Interestingly, from the solution of the KNCA equations at
$T = 0$ one also finds four different low-temperature
behaviors, corresponding to the precise four basins of
attraction listed above.

\section{Beyond the NCA}
\label{app:appC}

From the discussion in the preceding Appendix it is clear that
any proper theory of the Kondo effect must include particle-particle
and particle-hole processes on equal footing. Diagrammatically,
this necessitates going beyond the NCA class of diagrams, which
is the objective of the present Appendix.

\begin{figure}
\centerline{
\vbox{ \epsfxsize=62mm \epsfbox {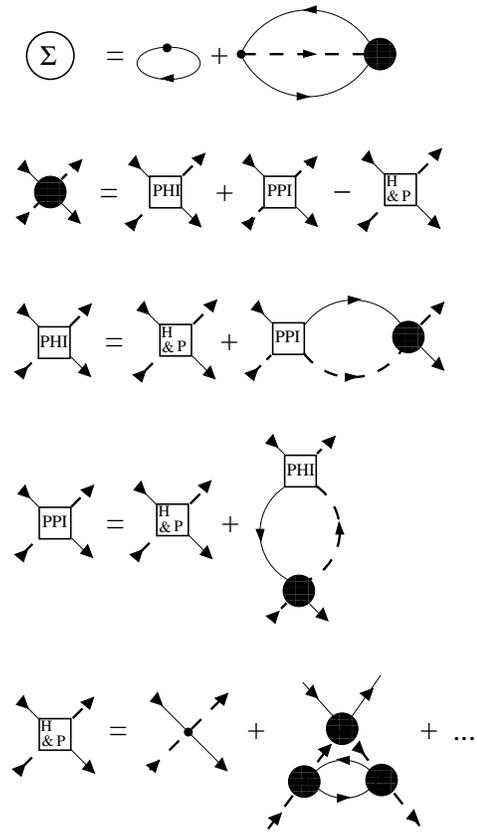}}
}\vspace{10pt}
\caption{A formally exact representation of the pseudo-fermion
         self-energy in terms of irreducible vertex functions.
         The first line represents all pseudo-fermion self-energy
	 diagrams. The full circle marks the full vertex function,
	 i.e., the collection of all connected four-legged diagrams
	 with one incoming and one outgoing pseudo-fermion line,
	 and one incoming and one outgoing conduction-electron
	 line. The full vertex function is expressed in turn in
	 the second line via the more elementary particle-hole
	 irreducible (PHI), particle-particle irreducible (PPI),
	 and both particle-hole and particle-particle irreducible
	 (PHI\&PPI) vertices. For precise definitions, see text.}
\label{fig:app8}
\end{figure}

Figure~\ref{fig:app8} displays a formally exact representation
of the pseudo-fermion self-energy. The first line in
Fig.~\ref{fig:app8} represents all pseudo-fermion self-energy
diagrams. The full circle marks the full vertex function, by
which we mean the collection of all connected four-legged
diagrams with one incoming and one outgoing pseudo-fermion
line, and one incoming and one outgoing conduction-electron
line. We classify the vertex diagrams according to two classes.
The particle-hole irreducible (PHI) vertex function is defined
as the sum of all vertex diagrams that cannot be disconnected
into two distinct vertex diagrams by cutting one internal
conduction-electron line and one internal pseudo-fermion
line propagating in {\em opposite} directions (third line of
Fig.~\ref{fig:app8}). The particle-particle irreducible (PPI)
vertex function is defined by a similar collection of diagrams,
that cannot be disconnected into two distinct vertex diagrams
by cutting one internal conduction-electron line and one
internal pseudo-fermion line propagating in the {\em same}
direction (fourth line of Fig.~\ref{fig:app8}). Note that
each vertex diagram is a member of at least one of the two
classes PHI and PPI. Some vertex diagrams, like the bare
vertex, are both PHI and PPI. We refer hereafter to these
joint diagrams as PHI\&PPI. Hence the full vertex function
is equal to the sum of the PHI and PPI vertex functions,
minus the PHI\&PPI diagrams which are counted twice.

There are infinitely many diagrams that are PHI\&PPI,
the first two of which are depicted on the fifth line
of Fig.~\ref{fig:app8}. Provided one is able
to sum the entire class of PHI\&PPI diagrams, then
Fig.~\ref{fig:app8} constitutes an exact representation of the
pseudo-fermion self-energy, as well as of the PHI, PPI, and
full vertex functions. In practice, it is impossible to sum
all the PHI\&PPI diagrams. However, any selection of a subclass
of PHI\&PPI diagrams that preserves
particle-hole symmetry defines an approximation scheme
that necessarily respects particle-hole symmetry. The
simplest choice for the PHI\&PPI vertex function is to
settle with the bare vertex, which defines the parquet
class of diagrams. Below we focus on this particular choice.

Each of the PHI, PPI and full vertex functions depend on three
energy variables, as well as on the spin indices of the four
external legs. Adopting the notation of Eq.~(\ref{V_abgd_1}),
we label the spin index of the outgoing (incoming)
conduction-electron line by $\alpha$ ($\alpha'$), and
that of the outgoing (incoming) pseudo-fermion line by
$\beta'$ ($\beta$). As for the energy variables, we
choose to work with the following combinations:
(i) The particle-hole energy, $z_{h}$, corresponding to
    the outgoing pseudo-fermion frequency minus the
    incoming conduction-electron frequency;
(ii) The particle-particle energy, $z_{p}$, corresponding
    to the incoming pseudo-fermion frequency plus the
    the incoming conduction-electron frequency; and
(iii) The pseudo-fermion energy, $z_f$, corresponding
    to the outgoing pseudo-fermion frequency.
For each of the PHI, PPI, and full vertex functions,
the projected functions are defined as
\begin{eqnarray}
&& \Gamma_{\alpha\beta;\alpha'\beta'}(z_{h},z_{p},z_f)
      =  \nonumber \\
&& \;\;\;\;\;\;\;\;\;\;\;\; \lim_{\lambda \to -\infty}
\Gamma^{(\lambda)}_{\alpha\beta;\alpha'\beta'}(z_{h} - \lambda,
      z_{p} - \lambda, z_f - \lambda) ,
\label{Gamma_project}
\end{eqnarray}
where $\Gamma^{(\lambda)}_{\alpha\beta;\alpha'\beta'}$ is
the corresponding unprojected function.

Approximating the PHI\&PPI vertex function by the
bare vertex, one can show that the projected PHI, PPI, and
full vertex functions have the form
\begin{eqnarray}
&& \Gamma^{(PHI)}_{\alpha\beta;\alpha'\beta'}
         (z_{h},z_{p},z_f) = \nonumber \\
&& \;\;\;\;\;\;\;\;\;\;\;\;\;\;
         \frac{1}{4} \sum_{\mu = 0, x, y, z}
         {\cal H}_{\mu}(z_{h},z_{p},z_f)
         \sigma^{\mu}_{\alpha\alpha'}
         \sigma^{\mu}_{\beta'\beta}, \\
&& \Gamma^{(PPI)}_{\alpha\beta;\alpha'\beta'}
         (z_{h},z_{p},z_f) = \nonumber \\
&& \;\;\;\;\;\;\;\;\;\;\;\;\;\;
         \frac{1}{4} \sum_{\mu = 0, x, y, z}
         {\cal P}_{\mu}(z_{h},z_{p},z_f)
         \sigma^{\mu}_{\alpha\alpha'}
         \sigma^{\mu}_{\beta'\beta}, \\
&& \Gamma^{(Full)}_{\alpha\beta;\alpha'\beta'}
         (z_{h},z_{p},z_f) = \nonumber \\
&& \;\;\;\;\;\;\;\;\;\;\;\;\;\;
         \frac{1}{4} \sum_{\mu = 0, x, y, z}
         {\cal J}_{\mu}(z_{h},z_{p},z_f)
         \sigma^{\mu}_{\alpha\alpha'}
         \sigma^{\mu}_{\beta'\beta} .
\end{eqnarray}
Thus, each of the above vertex functions retains the
form of the bare interaction, only with renormalized
couplings that are energy dependent. Furthermore, one
has the identity ${\cal J}_{\mu} = {\cal H}_{\mu} +
{\cal P}_{\mu} - J_{\mu}$, where $J_{\mu}$ are the
bare coupling constants.
The PHI and PPI vertex functions are given in turn
by the following set of coupled integral equations:
\begin{eqnarray}
{\cal P}_{0}(z_{h},z_{p},z_f) =&&
       J_{0} - \sum_{\mu = 0, x, y, z} [ {\cal H}_{\mu} \oplus
       {\cal J}_{\mu} ] , 
\label{BNCA_first} \\
{\cal P}_{i}(z_{h},z_{p},z_f) =&&
       J_{i} - [ {\cal H}_{0} \oplus {\cal J}_{i} ]
       - [ {\cal H}_{i} \oplus {\cal J}_{0} ]
\nonumber \\
&& - \sum_{j, m = x, y, z} \varepsilon_{ijm}^2
       [ {\cal H}_{j} \oplus {\cal J}_{m} ] , \\
{\cal H}_{0}(z_{h},z_{p},z_f) =&&
       J_{0} + \sum_{\mu = 0, x, y, z} [ {\cal P}_{\mu}
       \ominus {\cal J}_{\mu} ] , \\
{\cal H}_{i}(z_{h},z_{p},z_f) =&&
       J_{i} + [ {\cal P}_{0} \ominus {\cal J}_{i} ]
       + [ {\cal P}_{i} \ominus {\cal J}_{0} ]
\nonumber \\
&& - \sum_{j, m = x, y, z} \varepsilon_{ijm}^2
       [ {\cal P}_{j} \ominus {\cal J}_{m} ] ,
\end{eqnarray}
where $\epsilon_{ijm}$ is the antisymmetric tensor, $i$ is equal
to $x, y,$ or $z$, and the convolution operators $\oplus$ and
$\ominus$ stand for
\begin{eqnarray}
&& [{\cal H}_{\mu} \oplus {\cal J}_{\nu}] = \frac{1}{4}
         \sum_{k} f(\epsilon_k) G(z_h + \epsilon_k)
         {\cal H}_{\mu}(z_h, z_p + \Delta_1, z_f)
\nonumber \\
&& \;\;\;\;\;\;\;\;\;\;\;\;\;\;\;\;\;\;\;\;\;\;\;\;\;\;\;\;\;
         \times {\cal J}_{\nu}(z_h, z_p + \Delta_2,
         z_f + \Delta_2) ,
\label{conv_H_J} \\
&&[ {\cal P}_{\mu} \ominus {\cal J}_{\nu} ] = \frac{1}{4}
         \sum_{k} f(-\epsilon_k) G(z_p - \epsilon_k)
         {\cal J}_{\nu}(z_h - \Delta_2, z_p, z_f)
\nonumber \\
&& \;\;\;\;\;\;\;\;\;\;\;\;\;\;\;\;\;\;\;\;\;\;\;\;\;\;\;\;\;
         \times {\cal P}_{\mu}(z_h - \Delta_1, z_p,
         z_f - \Delta_1) .
\label{conv_P_J}
\end{eqnarray}
Here $\Delta_1$ and $\Delta_2$ are shorthands for
\begin{eqnarray}
\Delta_1 &=& \epsilon_k + z_f - z_p ,
\label{delta_1} \\
\Delta_2 &=& \epsilon_k + z_h - z_f .
\label{delta_2}
\end{eqnarray}
(Note that the frequencies $z_p - z_f$ and $z_f - z_h$ are
just the outgoing and incoming conduction-electron
frequencies, respectively.)
Omitting the constant term $( M J_0 N_e )/4$ ($N_e$
being the total number of conduction electrons), the
pseudo-fermion self-energy is equal to
\begin{eqnarray}
\Sigma(z) = && \frac{M}{8} \sum_{k,k'}
          f(\epsilon_k) f(-\epsilon_{k'})
          G(z + \epsilon_k - \epsilon_{k'}) \times
\nonumber \\
&& 
          \sum_{\mu = 0, x, y, z} J_{\mu}
          {\cal J}_{\mu}(z - \epsilon_{k'}, z + \epsilon_{k}, z) .
\label{BNCA_Sigma}
\end{eqnarray}

Equations~(\ref{BNCA_first})--(\ref{BNCA_Sigma}) represent
a complete summation of the parquet class of diagrams,
which includes in particular all diagrams contained within
poor-man's scaling. While we believe these equations are
intimately related to the conserving $T$-matrix approach
of Kroha {\em et al.}~\cite{CTMA} for the multi-channel
Anderson model, we are unable to formally link the two
approaches. We further emphasize that
Eqs.~(\ref{BNCA_first})--(\ref{BNCA_Sigma}) apply
to a general anisotropic Kondo model, ferromagnetic or
antiferromagnetic, and hence extend beyond
the Schrieffer-Wolff limit of the Anderson Hamiltonian.

A full self-consistent solution of
Eqs.~(\ref{BNCA_first})--(\ref{BNCA_Sigma}) is a difficult
task, due to the three energy variables entering the coupled
integral equations for ${\cal P}_{\mu}(z_h,z_p,z_f)$ and
${\cal H}_{\mu}(z_h,z_p,z_f)$. Below we resort to a crude
approximate treatment, aimed at demonstrating the amendment
of the KNCA flaws.

Since the onset the Kondo effect is manifest in the development
of strong scattering at the Fermi level, we set the energies
and thus frequencies of the incoming and outgoing conduction
electrons equal to zero. [We emphasize that, upon implementing the
projection of Eqs.~(\ref{Gamma_project}), the conduction-electron
frequencies acquire the value of the bare conduction-electron energies.]
We begin by noting that the energies $\Delta_1$ and $\Delta_2$
in each of Eqs.~(\ref{conv_H_J}) and (\ref{conv_P_J}) correspond
to the single-particle energy difference between the conduction
electrons incoming and outgoing one of the vertex function in that
equation --- ${\cal J}_{\nu}$ for $\Delta_1$, and ${\cal H}_{\mu}$
or ${\cal P}_{\mu}$ for $\Delta_2$.
Thus, since scattering in the Kondo problem is strong
near the Fermi level, one expects the main contribution to the $k$
summations in Eqs.~(\ref{conv_H_J})--(\ref{conv_P_J}) to come
from that range in $k$ where $\Delta_1$ and $\Delta_2$ are small.
As a first approximation, we therefore decouple the sums in
Eqs.~(\ref{conv_H_J})--(\ref{conv_P_J}) by setting
$\Delta_1 = \Delta_2 = 0$.

The second approximation we adopt is to linearize
Eqs.~(\ref{conv_H_J})--(\ref{conv_P_J}) by replacing the
PHI and PPI vertex functions, ${\cal H}_{\mu}$ and
${\cal P}_{\mu}$, with the bare vertex, $J_{\mu}$. This
clearly is a crude approximation, motivated primarily by
technical considerations. It encompasses, however, the ladder
summation of the KNCA. Using the above two approximations
in combination with the identity ${\cal J}_{\mu} = {\cal P}_{\mu}
+ {\cal H}_{\mu} - J_{\mu}$, we arrive at the following
set of equations for the full vertex functions
${\cal J}_{\mu} \equiv {\cal J}_{\mu}(z_f, z_f, z_f)$:
\begin{eqnarray}
{\cal J}_{0} = J_0 &+& \frac{1}{4}
            \left [ \Pi_{+}(z_f)\!-\!\Pi_{-}(z_f) \right ]
            \sum_{\mu = 0, x, y, z} J_{\mu} {\cal J}_{\mu} ,
\label{J_0_crude} \\
{\cal J}_i = J_i &+& \frac{1}{4}
            \left [ \Pi_{+}(z_f)\!-\!\Pi_{-}(z_f) \right ]
            \left [J_0 {\cal J}_i\!+\!J_i {\cal J}_{0} \right ]
\nonumber \\
&+& \frac{1}{4} \left [ \Pi_{+}(z_f)\!+\!\Pi_{-}(z_f) \right ]\!\!
            \sum_{j, m = x, y, z}\!\!\varepsilon_{ijm}^2 J_j {\cal J}_m .
\label{J_i_crude}
\end{eqnarray}
Here, as before, $i$ is equal to $x, y,$ or $z$, while
$\Pi_{\pm}(z)$ are the projected particle-hole and
particle-particle bubbles:
\begin{equation}
\Pi_{\pm}(z)= -\sum_{k} f(\pm \epsilon_k)
              G(z \pm \epsilon_k) .
\label{Bubbles}
\end{equation}

Similar to the KNCA, one can estimate the exponential dependence
of the Kondo temperature on $1/J_{\mu}$ by substituting the bare
pseudo-fermion Green function into Eq.~(\ref{Bubbles}), and
examining the location of the zero-temperature poles in the
full vertex functions ${\cal J}_{\mu}(z_f, z_f, z_f)$.
For a particle-hole symmetric band, when $\Pi_{+}(z)$ and
$\Pi_{-}(z)$ coincide, ${\cal J}_{0}(z_f, z_f, z_f)$ is
simply equal to $J_0$. Thus, there is no renormalization
of the potential scattering at the Fermi level, in
accordance with poor-man's scaling. In the isotropic case,
$J_x = J_y = J_z = J$, the remaining components of the full
vertex function are also isotropic, and are given by
\begin{equation}
{\cal J}(z_f, z_f, z_f) = \frac{J}{1 - J \Pi(z_f)} .
\end{equation}
Setting $T = 0$ and $z_f = \epsilon + i\delta$, one sees that
there are no divergences in ${\cal J}(z_f, z_f, z_f)$ for
$J < 0$, implying the absence of a ferromagnetic Kondo effect.
By contrast, ${\cal J}(z_f, z_f, z_f)$ diverges for $J > 0$
and $T = 0$ at $\epsilon \sim -D \exp [-1/\rho_0 J]$, which
we identify with $-k_B T_K$. Hence the correct exponential
dependence of $T_K$ on $1/J$ is properly recovered. For the
anisotropic Hamiltonian of Eq.~(\ref{H_matveev}), a similar
analysis gives $k_B T_K \sim D \exp [-\sqrt{2}/\rho_0 J]$,
which still deviates from the desired form of
$T_K \sim D \exp[-\pi/2\rho_0 J_{\perp}]$, but marks a
significant improvement over the KNCA.

Although the above treatment is too crude to convincingly
establish the amendment of the NCA flaws, we take it as
an encouraging indication for the elimination of the spurious
ferromagnetic Kondo effect and the correction of the
exponential dependence of $T_K$ on $1/J_{\mu}$ within
the proposed scheme.


\begin{thebibliography}{9}
\bibitem{comment_on_stability} Note that, while the intermediate-coupling
        fixed point of the multi-channel Kondo model with more than two
        conduction-electron channels is also stable against exchange
        anisotropy, the Wilson ratio is no longer universal as in the
        one- and two-channel cases (see Ref.~\onlinecite{Ye96}).
\bibitem{comment_on_exp} Note that, in the experiment of
        Ref.~\onlinecite{Ashoori99}, $e^2/C_B$ is equal to
        $0.46{\rm meV}$ while $T = 260{\rm mK}$,
        hence $e^2/C_B \approx 20 k_B T$.
\bibitem{comment_on_rho} In the general case where the density of
        states in the quantum box and in the lead are asymmetric
	functions of energy with different functional forms, the
	position of the degeneracy point will typically shift
	from $\phi = e(N + \frac{1}{2})/C_B$. This shift can be
	absorbed, though, into a redefinition of the ``magnetic
	field'' $eU$.
\bibitem{comment_on_chi_zz} More precisely, $\chi(H, T)$ is the
        $\chi_{zz}$ component of the isospin susceptibility tensor.
\bibitem{comment_on_sigma} For a spin-$\frac{1}{2}$ impurity, the
        pseudo-fermion Green function is diagonal in the spin index,
	for arbitrary $J_x, J_y$, and $J_z$. This is no longer
	the case if $J_x \neq J_y$ and the impurity spin is larger
	than one-half.
\bibitem{comment_on_constant_part} For convenience, we have omitted
        here a constant term, $\frac{M}{2}\sum_k J_0$, from the
        pseudo-fermion self-energy. The sole effect of such term is
        to shift the reference energy for the pseudo-fermion Green
        functions. It does not enter any physical observable.
\bibitem{comment_on_ferro} Actually, different threshold exponents
        apply to $J_z = -|J_{\perp}|$ and $J_z < -|J_{\perp}|$,
        both of which differ from the characteristic exponents for
        the $J_z > -|J_{\perp}|$ ``antiferromagnetic'' case.
\bibitem{comment_on_smearing} The shoulders of the charge steps
        are smeared over a larger energy scale, governed by
	$\pi \rho_0 t^2$.
\bibitem{comment_on_FL} More specifically, for $H \neq 0$ the
        physics is that of a Fermi gas interacting with a
        spin-polarized scattering center. For a detailed
        discussion, see Ref.~\onlinecite{CZ98},
        Secs. 3.1.2, 4.2e, 5.1.4, 6.1.2, and 7.2.2.
\bibitem{comment_on_e_0} For numerical convenience, we fixed
        $\epsilon_0$ at each iteration by
        requiring that the integrated spectral weights of
        $a_{\uparrow}(\epsilon)$ and
        $a_{\downarrow}(\epsilon)$ add up to one.
\bibitem{comment_on_CS} Similar to the NCA formulation for the
        Anderson model (see, e.g., Ref.~\onlinecite{Bickers87}),
        the magnetic vertex correction in the Coqblin-Schrieffer limit
        involves summation over an unpaired spin index that comes from
	the bare vertex. As a result, the magnetic
        vertex correction identically vanishes.
\bibitem{comment_on_w_c} The weak-coupling fixed point corresponds to a
        free spin. Therefore, the flow to weak coupling is signaled
	within the KNCA by the appearance of a zero-temperature delta
	peak in the pseudo-fermion Green function at the threshold
	energy, and the absence of a divergence in
	$\Delta(\epsilon + i\delta)$.
\bibitem{comment_on_A_m} Indeed, 
        $\rho_0 A_{\mu}/[1 - A_{\mu}\Pi(D + i\delta)]$ with
	$\Pi(D + i\delta)$ evaluated to leading logarithmic
	order is the solution of Eq.~(\ref{d_A_mu}).
\end{thebibliography}
\end{document}